\documentclass[preprint,amsmath,amssymb,prb,]{revtex4-2}

\usepackage{mathptmx}      
\usepackage{latexsym}
\usepackage{graphicx}
\usepackage{amssymb}
\usepackage{lineno}
\usepackage{amsmath}
\usepackage{times}
\usepackage{lipsum}
\usepackage{caption}
\usepackage{subcaption}
\usepackage{dcolumn}
\usepackage{bm}
\usepackage{color}
\usepackage{float}
\usepackage{stmaryrd}
\usepackage{xcolor}

\begin{document}

\title{Contribution of Mach number on the evolution of Richtmyer-Meshkov instability induced by shock-accelerated square light bubble}
		
\author{Satyvir Singh}
\email{satyvir.singh@ntu.edu.sg}
\affiliation{School of Physical and Mathematical Sciences,
Nanyang Technological University, 21 Nanyang Link, Singapore – 637371}
	

\begin{abstract}

The Richtmyer-Meshkov (RM) instability has long been an interesting subject due to its fundamental significance in scientific research, as well as its crucial role in engineering applications. In this study, the contribution of shock Mach number on the evolution of the RM instability induced by a shock-accelerated square light bubble is investigated numerically. The square bubble is composed of helium gas and the surrounding (ambient) gas is nitrogen. Three cases of incident shock strength are considered: $\text{M}_{s}=$1.21, 1.7, and 2.1. An explicit mixed-type modal discontinuous Galerkin scheme with uniform meshes is employed to numerically solve a two-dimensional system of unsteady compressible Navier--Stokes--Fourier equations. The numerical results show that the shock Mach number plays an important role during the interaction between a planar shock wave and a square light bubble. The shock Mach number causes significant changes in flow morphology, resulting in complex wave patterns, vorticity generation, vortex formation, and bubble deformation. In contrast to low Mach numbers, high Mach numbers produce the larger rolled-up vortex chains, larger inward jet formation, and a stronger mixing zone with greater expansion. The effects of Mach numbers are explored in detail through phenomena such as the vorticity generation, and evolutions of enstrophy as well as dissipation rate. Finally, the Mach number effects on the time-variations of the shock trajectories and interface features are comprehensively analyzed.
\end{abstract}	

\maketitle

\section{Introduction}
\label{Sec.1}

The Richtmyer-Meshkov (RM) instability is a fundamental physical phenomenon that occurs when a moving shock wave passes through a gas interface separating fluids of different densities. Small perturbations appear on the interface in the form of light bubbles floating into the heavy fluid and heavy spikes penetrating into the light fluid due to a misalignment of the pressure gradient $(\nabla p)$ across the shock and the local density gradient $(\nabla \rho)$ at the interface.
These perturbations become greater over time, and more complex structures arise, generating a mixing zone of light and heavy fluids that interpenetrate one another.
Richtmyer \cite{Richtmyer1960taylor} was the first to investigate this shock-induced instability analytically and numerically, and Meshkov \cite{Meshkov1969Instability} later validated it in a shock tube. This Instability is considered as an impulsive or shock-accelerated version of the continuously driven Rayleigh--Taylor instability \cite{taylor1950instability} that develops at accelerated density-stratified interfaces. The RM instability is ubiquitous in a wide range of natural and engineering applications, including supernova explosions, lithotripsy, astrophysics, inertial confinement fusion, scramjet combustion process, and many more. Over the last several decades, numerous studies on the RM instability have been conducted, and extensive reviews of the occurrence and applications of this instability have been presented by Zabusky \cite{zabusky1999vortex}, Holmes \cite{holmes1999richtmyer}, Brouillette \cite{brouillette2002richtmyer}, Ranjan et al. \cite{ranjan2011shock}, Luo et al. \cite{luo2014experimental}, and Zhou \cite{zhou2017rayleighI,zhou2017rayleighII}.

The study of shock-accelerated bubble has been one of the most fundamental research topics for characterizing the physical configuration of the RM instability over the years.
The shock-accelerated bubble was initially explored experimentally by Markstein \cite{markstein1957shock}, Rudinger, and Somers \cite{rudinger1960behaviour} in their groundbreaking research works. Haas and Sturtevant \cite{haas1987interaction} studied experimentally the interaction of a plane weak shock wave with a single gas bubble containing either helium or R22 gas. In their experiments, shadowgraph photography was used to visualize the evolving waves and the distortion of the bubble as a result of the incoming shock. Layes et al. \cite{layes2005experimental, layes2009experimental} experimentally investigated the shock-accelerated bubbles using high-speed shadowgraphy, and a vortex ring and inward jet were observed for different gases. Kumar et al. \cite{kumar2005stretching} conducted experimental studies of the acceleration of different configurations in shock-accelerated cylinder heavy bubbles at $\text{M}_{s}=1.2$ with the aid of planar laser--induced fluorescence technique, and showed that the early-time stretching rate of material lines after the interaction depending on the configuration and orientation of the gaseous bubbles. Ranjan et al. \cite{ranjan2007experimental} investigated experimentally the compression and unstable evolution of a shock-accelerated spherical helium bubble at shock Mach number $(\text{M}_{s}=2.95)$ in a vertical shock tube, and explored the different vortex rings of the distorted bubble. Following that, Ranjan et al. \cite{ranjan2008shock} studied the divergent--geometry shock-accelerated bubble in the shock Mach number range of $1.4 \leq \text{M}_{s} \leq 3.0$, and the experimental planar laser diagnostics resolved features such as the formation of a long-lived primary vortex ring, as well as counter--rotating secondary and tertiary upstream vortex rings, which appeared for high shock Mach numbers. Haehn et al. \cite{haehn2010experimental,haehn2012experimental} explored experimentally the shock-accelerated spherical bubble under re-shock situations with three shock Mach numbers $(\text{M}_{s}=1.35-2.33)$ to observe the evolution of the vortex ring. Si et al. \cite{si2012experimental} investigated experimentally the evolution of spherical light/heavy gas bubbles in flows accelerated by incident and reflected shocks using high-speed schlieren photography, as well as the impact of reflected distances on on the flow morphology.

To better understand the physical causes of the RM instability, numerous computational studies for shock-accelerated bubbles have also been conducted.
Comprehensive studies on the shock-accelerated bubbles were performed by Picone and Boris \cite{picone1988vorticity} and Quirk and Karni \cite{quirk1996dynamics}, and the experimental results of Haas and Sturtevant \cite{haas1987interaction} were reproduced. 
Zabusky and Zeng \cite{zabusky1998shock} simulated planar shocks interacting with an $\text{R}_{12}$ axisymmetric spherical bubble, and observed that the collapsing shock cavity within the bubble caused an expelled weak jet at low Mach number, but at higher Mach numbers (e.g. $\text{M}_{s}=2.5$), the ‘vortical projectiles’ appear on the downstream side of the bubble. Bagabir and Drikakis \cite{bagabir2001mach} numerically examined the effects of the incident shock Mach numbers $(\text{M}_{s}=1.22-6.0)$ on the flow evolution of the shock-accelerated light bubbles, and explored additional gas dynamic features as the Mach number increased. Giordano and Burtschell \cite{giordano2006richtmyer} investigated the RM instabilities by analyzing of shocks-bubble interaction at low Mach number $(\text{M}_{s}=1.2)$ for understanding the vortex deposition by the baroclinic terms. Niederhaus et al. \cite{Niederhaus2008} performed three-dimensional multifluid Eulerian simulations to investigate the flow morphologies and integral properties of shock-accelerated bubbles, and explored the contributions of different Atwood numbers $(-0.8 < A_{t} < 0.7)$ and shock intensities $(1.1 \leq \text{M}_{s} \leq 5.0)$. Using the high-resolution computation scheme, Zhu et al. \cite{zhu2018numerical} numerically explored the impacts of different incident shock Mach numbers $(1.21 \leq \text{M}_{s} \leq 2.1)$ on the flow fields of a shock-accelerated $\text{SF}_{6}$ bubble, and observed that the bubble distorts separately with the increasing incident shock Mach number. At high Mach numbers $(\text{M}_{s}=2.5-3.0)$, Rybakin and Goryachev \cite{rybakin2014supersonic} studied numerically the deformation and instability of a low-density gas bubble, the formation and evolution of vortex rings, and the shock wave--bubble configuration. Recently, Singh and Battiato \cite{singh2021nonequilibrium} analyzed numerically the behavior of a shock-accelerated cylindrical heavy bubble at low Mach number $(\text{M}_{s}=1.21)$ under the nonequilibrium conditions of diatomic and polyatomic gases. Further, this research work was extended by Singh et al. \cite{singh2021bulk} to investigate the impact of bulk viscosity on the flow morphology of a shock-accelerated cylindrical light bubble. It was observed that the  bulk viscosity of a gas molecule, which is directly related to its rotational mode, plays a vital role in the the interaction process.

Most of existing experimental, theoretical, and numerical studies have focused on shock-accelerated bubbles of various shapes, such as cylindrical, elliptical, or spherical. A few studies have been conducted on the shock-accelerated bubbles with polygonal interface shapes that contribute promising conditions for the shock refraction physical phenomena. The interaction of a planar shock wave with light polygonal interfaces (square, triangle, and diamond) in the slow/fast configuration was studied experimentally and numerically by Zhai et al. \cite{zhai2014interaction}. 
Luo et al. \cite{luo2015interaction} then investigated experimentally the shock refraction phenomenon at a fast/slow interface, as well as the contributions of the initial interface shape on the RM instability by interacting a planar shock wave with six different heavy polygons (a square, two rectangles, two triangles, and a diamond). Igra and Igra \cite{igra2018numerical} studied numerically the interaction of a planar shock wave with square and triangular bubbles containing different gases, inspired by the research works of Zhai and co-authors \cite{zhai2014interaction,luo2015interaction}. The interaction of a planar shock wave propagating in air with a polygonal bubble (composed of two triangles) containing two different gases was then investigated in the work of Igra and Igra \cite{igra2020shock}. Various simulations based on heavy gas inhomogeneities with some simple geometries (square, rectangle, circle, and triangle) was performed numerically by Fan et al. \cite{fan2019numerical} to determine the source of the jet formation. Singh \cite{singh2020role} investigated numerically the impacts of the Atwood numbers on the flow evolution of a shock-accelerated square bubble containing various gases at low Mach number $(\text{M}_{s}=1.22)$, and observed that the Atwood number has a significant impact on the flow evolution with complex wave pattern, vortex creation, vorticity generation, and bubble deformation. Recently, Singh \cite{singh2021IJHMT} explored numerically the thermal non-equilibrium effects of diatomic and polyatomic gases on the flow dynamics of a shock-accelerated square light bubble.

The strength of the incident shock wave is well known as a critical controlling parameter for the investigation of compressible flows such as, compressible hydrodynamic instability, 
and turbulence mixing. Therefore, the effects of different Mach numbers ($\text{M}_{s}=1.21,1.7$, and 2.1, which represent the weak, intermediate, and strong shock cases, respectively) on the evolution of the RM instability in a shock-accelerated square light bubble are examined numerically in the current study. To the best of the authors’ knowledge, there have been no reports in the literature on the effect of Mach number on the evolution of the RM instability in a shock-accelerated square light bubble. Utilizing numerical simulations based on an explicit mixed-type modal discontinuous Galerkin method, the effects of Mach numbers on the wave patterns, bubble deformation, vortex creation, vorticity generation, evolution of enstrophy, and dissipation rate, and interface features are discussed. The remainder of this paper is organized as follows: Sec. \ref{Sec.2} outlines the computational model including the problem setup and employed numerical method. Sec. \ref{Sec.3} presents the grid refinement analysis and the validation of the numerical method. Sec. \ref{Sec.4} discusses in detail the Mach number effect on the shock-accelerated square light bubble in terms of flow evolution, vorticity generation and their quantitative analysis. Sec. \ref{Sec.5} draws some concluding remarks with further development in this topic.

\section{Problem setup and Computational model}
\label{Sec.2}

\subsection{Problem setup}
\label{Sec.2.1}

\begin{figure}
	\centering
	\includegraphics[width=1.0\linewidth]{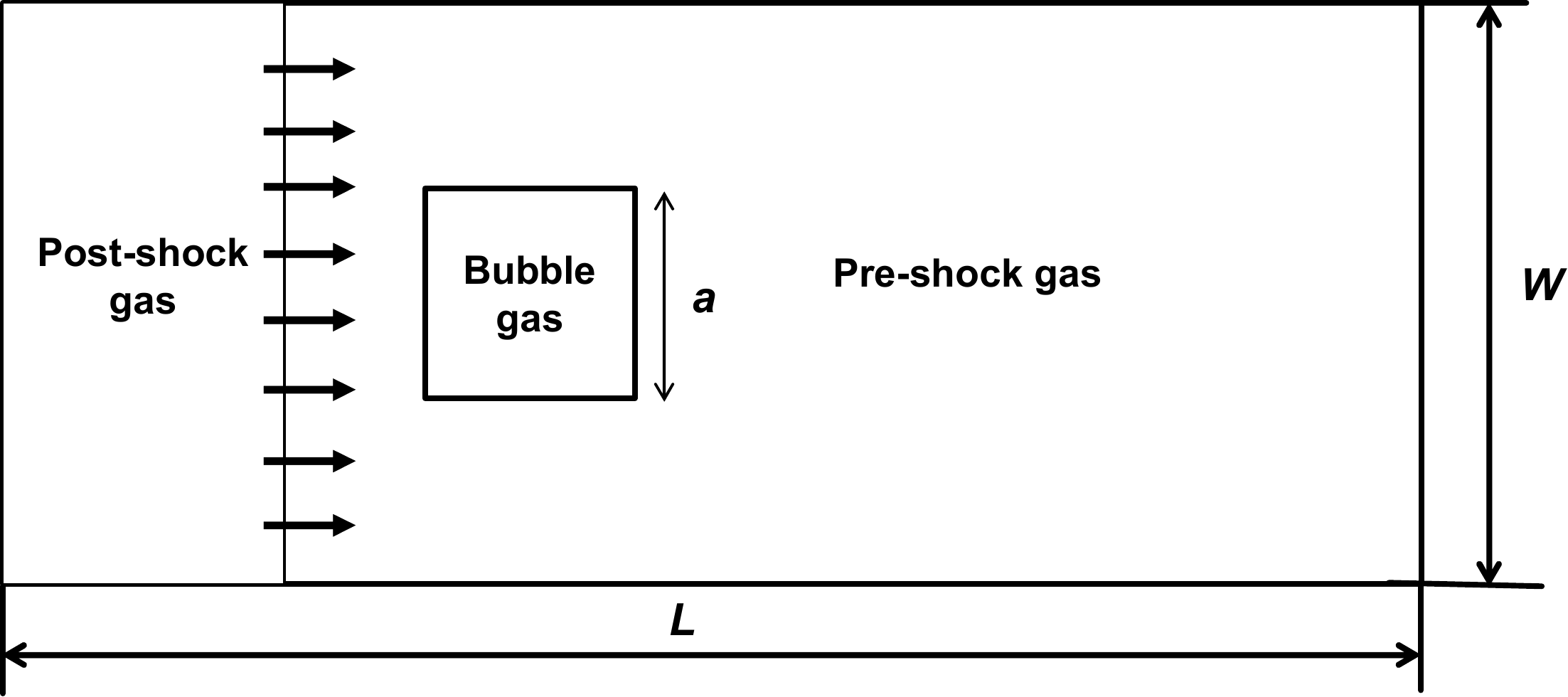}
	\caption{Schematic diagram of initial flow field and computational domain.}
	\label{Fig:1} 
\end{figure}

Figure \ref{Fig:1} illustrates a schematic diagram of the flow model used to simulate a shock-accelerated square light bubble surrounded with ambient gas. A rectangular domain of $[0,250]~\text{cm} \times [0,90]~\text{cm}$ is assumed for the numerical simulation of the shock-accelerated bubble problem, where a moving incident shock (IS) wave and a square stationary bubble are considered. In the computational domain, the IS wave with Mach number $\text{M}_{s}$ propagates from left to right, the membranes separating the two fluids rupture and the RM instabilities are generated on the bubble interface. The initial position of the shock wave is set to $x=30$ cm, from the left-hand side of the computational domain. The edge length of the square bubble is set to $a=40$ cm. The initial pressure and temperature are considered as $P_{0}=101,325$ Pa, and $T_{0}=273$ K, respectively around the square bubble. As helium gas has been widely adopted as a light gas in previous studies on the RM instability, we also consider helium gas with a density of $\rho_{b}=0.160 \times 10^{-3}$ $\text{g}/ \text{cm}^{3}$ inside the square bubble. Nitrogen gas is considered as an ambient gas with a density of $\rho_{g}=1.25 \times 10^{-3}$ $\text{g}/ \text{cm}^{3}$.

\subsection{Governing equations}
\label{Sec.2.2}

Basically, the compressible multispecies flow model is simulated with a gas mixture to solve shock-accelerated bubble problems \cite{giordano2006richtmyer,Shankar2011Kawai}. Interestingly, Picone and Boris, \cite{Picone1988} Samtaney and Zabusky, \cite{Samtaney1994} Quirk and Karni, \cite{quirk1996dynamics} and Bagabir and Drikakis \cite{bagabir2001mach} have found that assigning different specific heat capacities $\gamma$ to each gas does not affect the qualitative details of the vorticity generation, particularly the creation of large-scale structures. According to Quirk and Karni, \cite{quirk1996dynamics} for the problem of a shock-accelerated bubble ``...the errors introduced by the single-gas model assumption are not catastrophic and to some extent may be tolerated.'' However, such errors cannot be tolerated in applications such as air--fuel mixing in a supersonic combustion system, where temperature changes will substantially affect the mixing. Recently, Latini and Schilling \cite{Latini2020RMI} numerically investigated the growth dynamics of two- and three-dimensional single-mode reshocked air/$\text{SF}_{6}$ RM instabilities by considering a single specific heat ratio $\gamma$. Therefore, the present shock-accelerated  square light bubble problem is setup as an unsteady compressible laminar flow that assumes a single-component perfect gas with a specific heat ratio of $\gamma$.

Two-dimensional compressible Navier--Stokes--Fourier equations equations for the laminar flow model are considered here, which are written in conservation form as \cite{Xiao2014Myong,Prince2018,singh2020role}   
\begin{equation}
	\label{Eq:1}
	\frac{\partial \mathbf{U}}{\partial t} + \frac{\partial \mathbf{F}^{inv}}{\partial x} + \frac{\partial \mathbf{G}^{inv}}{\partial y} + \frac{\partial \mathbf{F}^{vis}}{\partial x} + \frac{\partial \mathbf{G}^{vis}}{\partial y}= 0,
\end{equation}
where
\begin{align*}
	\mathbf{U} &= \begin{bmatrix}  \rho , \rho u  , \rho v , \rho E  \end{bmatrix}, \\
	\mathbf{F}^{inv} &= \begin{bmatrix}  \rho u , \rho u^{2} +  p , \rho uv , (\rho E + p) u \end{bmatrix}, \\
	\mathbf{G}^{inv} &= \begin{bmatrix}  \rho v , \rho uv , \rho v^{2} +  p  , (\rho E + p) v \end{bmatrix}, \\
	\mathbf{F}^{vis} &= \begin{bmatrix}  0 , \Pi_{xx} , \Pi_{xy} , \Pi_{xx}u + \Pi_{xy}v + Q_{x}  \end{bmatrix}, \\
	\mathbf{G}^{vis} &= \begin{bmatrix}  0 , \Pi_{xy} , \Pi_{yy}  , \Pi_{xy}u + \Pi_{yy}v + Q_{y} \end{bmatrix}.
\end{align*}
Here, $\rho$ is the mass density, $u$ and $v$ are the velocity components in $x-$ and $y-$ directions, respectively. $E$ is the total energy density and $p$ is the static pressure determined by the ideal gas law as 
\begin{equation}
	\label{Eq:2}
	p = (\gamma-1)(\rho E - \frac{1}{2}(u^{2}+v^{2})),
\end{equation}
where $\gamma$ is the specific heat ratio. The value of $\gamma$ is considered to be $1.4$ for nitrogen gas. The symbols $\Pi_{xx}$, $\Pi_{xy}$ and $\Pi_{yy}$ are the components of shear stress vector $\Pi$ defined as follows 
\begin{align*}
	\Pi_{xx} & = - \mu \left[\frac{4}{3} \frac{\partial u}{\partial x} - \frac{2}{3} \frac{\partial v}{\partial y} \right], \\
	\Pi_{xy} & = - \mu \left[\frac{\partial v}{\partial x} + \frac{\partial u}{\partial y} \right], \\
	\Pi_{yy} & = - \mu \left[\frac{4}{3} \frac{\partial v}{\partial y} - \frac{2}{3} \frac{\partial u}{\partial x} \right].
\end{align*}
The symbols $Q_x$ and $Q_y$ are the heat fluxes in the $x-$ and $y-$ directions, respectively defined as
\begin{align*}
	Q_{x} = -\kappa \frac{\partial T}{\partial x}, \quad
	Q_{y} = -\kappa \frac{\partial T}{\partial y},  \\
\end{align*}
where $T$ is the absolute temperature. In the above expressions, the symbols $\mu$ and $\kappa$
represent the Chapman-Enskog shear viscosity, and the thermal conductivity, respectively. These expressions for the Chapman-Enskog linear transport coefficients can be employed as \cite{Xiao2014Myong,singh2020role}
\begin{equation}
	\label{Eq:3}
	\mu = \left(\frac{T}{T_{ref}}\right)^s,  \quad \quad \kappa = \left(\frac{T}{T_{ref}}\right)^s,
\end{equation}
where $T_{ref}$ is the reference temperature. Here, the value of $T_{ref}$ is considered as $T_{ref}=273.15$ K.
The symbol $s$ stands for the index of the inverse power laws of gas molecules, which is given as
\begin{equation}
	\label{Eq:4}
	s = \frac{1}{2} + \frac{2}{\nu-1}.
\end{equation}
Here, the parameter $\nu$ is the exponent of the inverse power laws for the gas particle interaction potentials. The value of $s$ is assumed to be 0.78 for nitrogen gas \cite{Bird194}.

\subsection{Initialization of the problem}
\label{Sec:2.3}

The evolution of the RM instability induced by a shock-interface interaction is very sensitive to the initial conditions, where the flow fields are dominated due to the baroclinic vorticity deposited on the interface. 
The baroclinic vorticity is directly proportional to the gradient of the initial density profile in the light gas bubble, hence necessitating accurate modeling of initial conditions.	Numerous experimental, theoretical and numerical studies have focused on the influence of initial conditions on growth of the RM instabilities. Hahn at al. \cite{hahn2011richtmyer} investigated the growth of RM instability and turbulent mixing under realistic conditions for the surface perturbations, including reshocked flow, through an inclined material interface with perturbations with different spectra but the the same standard deviation. Thornber et al. \cite{thornber2010influence} studied the effects of different three-dimensional multi-mode initial conditions on the rate of growth of a mixing layer initiated via a RM instability. Balasubramanian et al. \cite{balasubramanian2012experimental} analyzed experimentally the dependence of initial condition parameters, the amplitude $\delta$ and wavenumber $\kappa$ of perturbations, on turbulence and mixing in shock-accelerated RM unstable fluid layers. Recently, Mansoor et al. \cite{mansoor2020effect} investigated numerically the effect of initial conditions on the late-time growth and mixing transition of RM instability from sinuous perturbations on an air/$\text{SF}_{6}$ interface subjected to a weak planar shock wave.

In present study, an ambient condition on the right-hand side of the shock wave is employed to initialize the computational simulation for shock-accelerated bubble. The primitive variables are calculated on the left-hand side of the shock wave using the standard Rankine--Hugoniot conditions \cite{kundu2019high}. The standard Rankine--Hugoniot conditions for primitive variable calculations are expressed as
\begin{equation}
	\label{Eq:5}
	\begin{aligned}
		M_{2}^{2} & = \frac{1+\left[\frac{(\gamma-1)}{2}\right]\text{M}_{s}^2}{\gamma \text{M}_{s}^2 - \frac{(\gamma-1)}{2}}, \quad
		\frac{p_{2}}{p_{1}} & = \frac{1+\gamma \text{M}_{s}^2}{1+\gamma {M}_{2}^2}, \quad
		\frac{\rho_{2}}{\rho_{1}} & = \frac{\gamma -1 + (\gamma+1)\frac{p_2}{p_1}}{\gamma +1 + (\gamma-1)\frac{p_2}{p_1}}.\\
	\end{aligned}
\end{equation}
In the above expressions, $\text{M}_{s}$ denotes the shock Mach number, and the subscripts 1 and 2 denote the left- and right-hand sides of the shock wave, respectively. In present study, three different incident shock Mach numbers ($\text{M}_{s}=1.21, 1.7$, and 2.1) are selected for numerical simulations. As for the computational specifications, the left boundary is set to inlet, while the upper, bottom, and right boundaries are considered as outflow boundaries.

\subsection{Numerical method based on explicit modal discontinuous Galerkin scheme}
\label{Sec.2.4}

In order to obtain reliable quantitative predictions about the RM instability, the choice of the numerical method turns out to be absolutely crucial. Due to the nature of the problem, which involves the propagation of strong shocks, the so-called Godunov methods, which are based on the conservative formulation of the equations, should be preferred with respect to any other numerical method. Moreover, the numerical modeling of complex flow structures and fluid instabilities would also benefit significantly if high-resolution schemes are adopted. Several high-resolution schemes can be used in conjunction with finite difference, finite volume and spectral methods, as well as with adaptive grid refinement in block-structured and unstructured grid frameworks \cite{harten1997high,karniadakis2013spectral}. The performance of several high-resolution schemes in various unsteady, inviscid, compressible flows were investigated by Bagabir and Drikakis \cite{bagabir2004numerical}. Later, Mosedale and Drikakis \cite{mosedale2007assessment} employed the high-resolution and very high-order methods
for implicit large-eddy simulation to simulate the multispecies
two-dimensional single-mode RM instability problems.

Discontinuous Galerkin (DG) approaches have recently gained prominence in fields ranging from fluid mechanics to acoustics and electromagnetics \cite{Cockburn1998Local,Cockburn1998DGV,Xiao2014Myong,Prince2018,singh2017JCFE,Signh2018Thesis,Singh2020IACM,Singh2020Material,Chourushi2020Singh,singh2018non,singh2020topology,Singh2021CF,Singh2021JCP}. These methods are locally conservative, stable, and high-order accurate methods which can easily handle complex geometries, irregular meshes with hanging nodes, and approximations that have polynomials of different degrees in different elements.
In this paper, the two-dimensional compressible avier--Stokes--Fourier equations (\ref{Eq:1}) are solved by an in-house developed explicit mixed-type modal DG solver based on structured meshes \cite{singh2017JCFE,singh2021IJHMT,singh2018non,singh2020role}. The computational domain is discretized into rectangular elements, and scaled Legendre polynomial functions are employed for the elements. The Gauss--Legendre quadrature rule is implemented for both the volume and the boundary integrations, and the Roe flux \cite{Roe1981JCP} is applied for the inviscid term. The local DG scheme \cite{Cockburn1998Local} is employed for the auxiliary and viscous fluxes at the elemental interfaces. A polynomial expansion of third-order accuracy is used to approximate the solutions in the finite element space, and an explicit third-order accurate strong stability preserving Runge--Kutta scheme is used for the time integration. The nonlinear total variation bounded limiter proposed by Cockburn and Shu \cite{Cockburn1998DGV} is used to eliminate spurious numerical fluctuations in the solutions. 

\section{Grid refinement and validation study}
\label{Sec.3}

The most fascinating phenomenon is the visualization of flow evolution during the interaction process in the shock-accelerated interface problems. To visualize the computational results, numerical schlieren images based on  the magnitude of the gradient of the density field, which is defined as \cite{schilling2007physics,Marquina2003Mulet} 
\begin{equation}
	\label{Eq:6}
	S_{i,j} = exp \left(-k(\phi_{i,j}) \frac{|\nabla \rho_{i,j}|}{max_{i,j}|\nabla \rho_{i,j}|}\right),
\end{equation}
where
\begin{equation*}
	k(\phi_{i,j}) = \begin{cases}
		20 \quad \text{if} \; \phi_{i,j} >0.25, \\ 
		100 \quad \text{if} \; \phi_{i,j} <0.25.
	\end{cases}
\end{equation*}
For the numerical simulation, the computational time is considered as the non-dimensionalized to produce a dimensionless time scale $(\tau)$ defined as  
\begin{equation}
	\label{Eq:7}
	\tau = t \cdot \frac{c \, \text{M}_{s}}{a},
\end{equation}
where $t$ is the real computational time, $c$ is the local sound speed, $\text{M}_{s}$ is the Mach number of incident shock wave, and $a$ is the edge length of the square.

\subsection{Grid refinement analysis}
\label{Sec:3.1}

\begin{figure} [hbt!]
	\centering
	\includegraphics[width=0.7\linewidth]{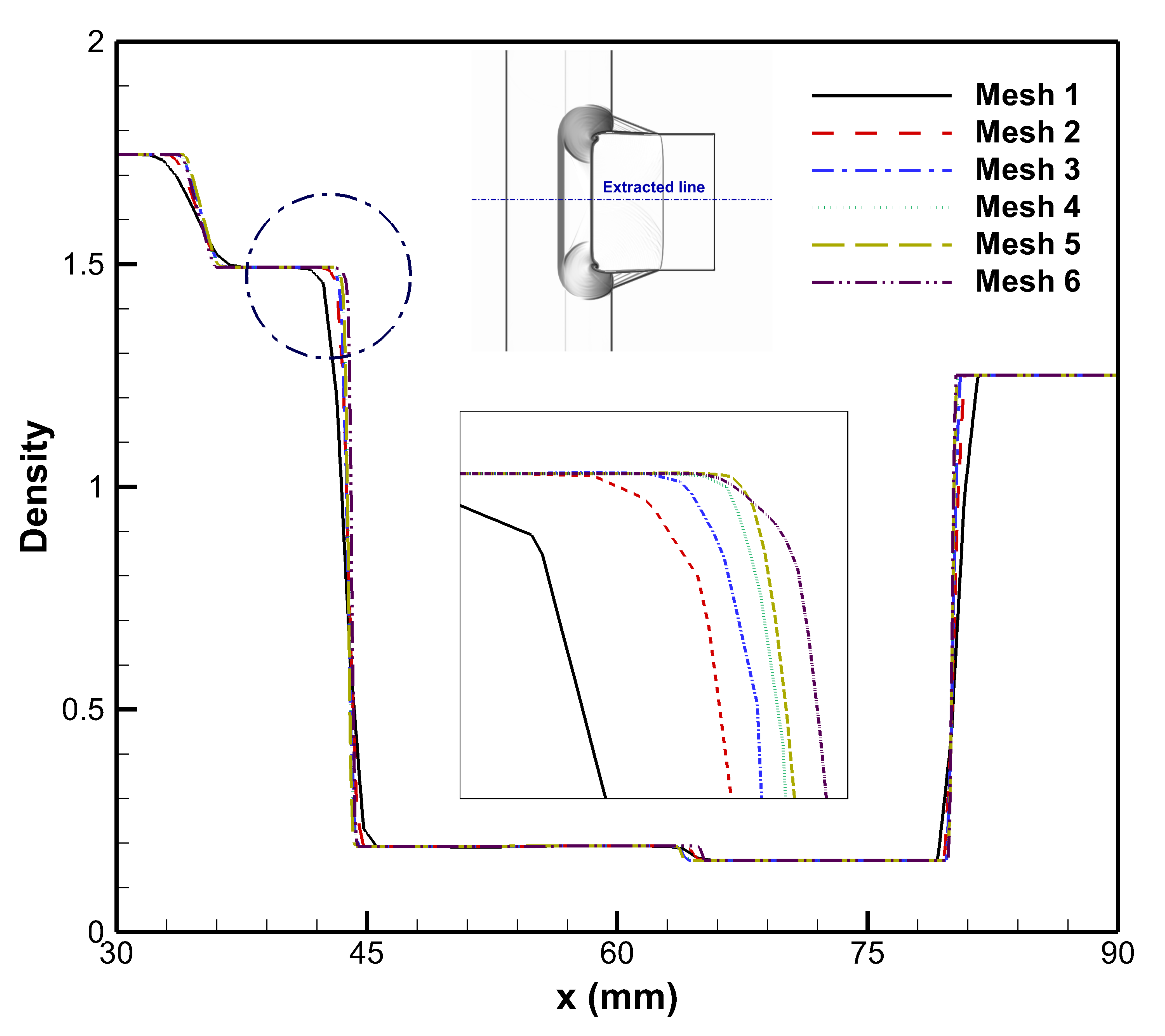}
	\caption{Grid refinement study in a shock-accelerated square helium bubble surrounded by nitrogen gas: density distribution profiles for different mesh sizes.}
	\label{Fig:2}
\end{figure}

A grid sensitive analysis is performed by computing one test case on a shock-accelerated square helium bubble surrounded by nitrogen gas at $\text{M}_{s}=1.21$ to accurately capture the complex structure of the flow field and the evolution process of the interface. For this purpose,
six uniform rectangular meshes are considered. The labels `Mesh 1'$-$ `Mesh 6' correspond to mesh points $200 \times 100$, $400 \times 200$, $600 \times 300$, $800 \times 400$, $1000 \times 500$ and $1200 \times 600$, respectively. When an incident shock wave hits the square bubble surface, the volume of the bubble is evidently compressed, and the bubble inside generated shock waves form an divergent shape. Figure \ref{Fig:2} depicts the profiles of the density distribution extracted with the center-line of the computed bubble at time $t=3$ to demonstrate the grid sensitivity. The results show that the Mesh 6 is very close to a asymptotic range. Based on this analysis, all subsequent computations are carried out using Mesh 6.

\subsection{Error estimation}
\label{Sec:3.2}

It is critical to determine the precision and accumulation of errors while doing large-scale simulations of complex combustion gas dynamics in unsteady-state flows.
The error generally depends on accuracy of numerical scheme and grid resolution, and on the number of time steps. For such problems, a numerical method to estimate the error accumulation and simulation precision was proposed by Smirnov et al. \cite{smirnov2014hydrogen, smirnov2015accumulation}.

In the one-dimensional case $S_{1}$, the relative error of integration is proportional to the mean ratio of the cell size $\Delta L$ to the domain size $L_{1}$ in the direction of integration in the power, and depends on the accuracy of the scheme:
\begin{equation}
	\label{Eq:8}
	S_{1} \equiv \left( \frac{\Delta L}{L_{1}} \right)^{k+1}.
\end{equation}
For a uniform grid, $S_{1} \equiv (1/N_{1})^{k+1}$, where $N_{1}$ is the number of cells in the direction of integration and $k$ is the order of accuracy of the numerical scheme. The errors provided by Eq. (\ref{Eq:8}) in two directions are being summed up:
\begin{equation}
	\label{Eq:9}
	S_{err} \equiv \sum_{i=1}^{2} S_{i}.
\end{equation}
The allowable value of the total error $S_{max}$ is typically 1--5\%, because the initial and boundary conditions are usually not known with a higher degree of accuracy. As a result, the following inequality should be satisfied:
\begin{equation}
	\label{Eq:10}
	S_{err} \cdot \sqrt{n} \leq S_{max},
\end{equation}
where $n$ is the number of time steps. The maximal allowable number of time steps can then be determined by the following formula:
\begin{equation}
	\label{Eq:11}
	n_{max} = \left( \frac{S_{max}}{S_{err}} \right)^{2}.
\end{equation}
and the reliability of results can be defined as
\begin{equation}
	\label{Eq:12}
	R_{s} = \frac{n_{max}}{n}.
\end{equation}

\begin{table}
	\caption{Error estimation.}
		\label{tab:1} 
		\begin{ruledtabular}
			\begin{tabular}{ccccccc}
				Allowable & Grid & Time & Number of & Accumulated & Allowable number & Reliability \\
				error $(\%)$ & resolution & simulated & time steps & error & of time steps & $(R_{s}=n_{max}/n)$ \\ 
				\hline
				5 & $200 \times 100$ & 4 & 83 & $9.32 \times 10^{-4}$ & 2878 & 35 \\
				5 & $400 \times 200$ & 4 & 179 & $3.45 \times 10^{-5}$ & $2.10 \times 10^{6}$ & $1.17 \times 10^{4}$ \\
				5 & $600 \times 300$ & 4 & 275 & $1.91 \times 10^{-5}$ & $6.85 \times 10^{6}$ & $2.49 \times 10^{4}$ \\
				5 & $800 \times 400$ & 4 & 371 & $1.21 \times 10^{-5}$ & $1.71 \times 10^{7}$ & $4.61 \times 10^{4}$ \\
				5 & $1000 \times 500$ & 4 & 466 & $9.05 \times 10^{-6}$ & $3.05 \times 10^{7}$ & $6.55 \times 10^{4}$ \\
				5 & $1200 \times 600$ & 4 & 561 & $6.93 \times 10^{-6}$ & $5.21 \times 10^{7}$ & $9.29 \times 10^{4}$ \\
			\end{tabular}
	\end{ruledtabular}
\end{table}

Table \ref{tab:1} predicts the accumulation of errors for the present DG scheme with different grid resolutions. The allowable error is considered to be 5\%, and the final simulation time is set to 4. As can be seen, the errors accumulate rapidly for the coarse grid and decrease as the grid resolution increases. The reliability of the results increases with a higher grid resolution and scheme accuracy. For the present simulations, all results demonstrate that the computational model is highly reliable, but this may not be the case for longer simulation periods.

\begin{figure} [hbt!]
	\centering
	\includegraphics[width=1.0\linewidth]{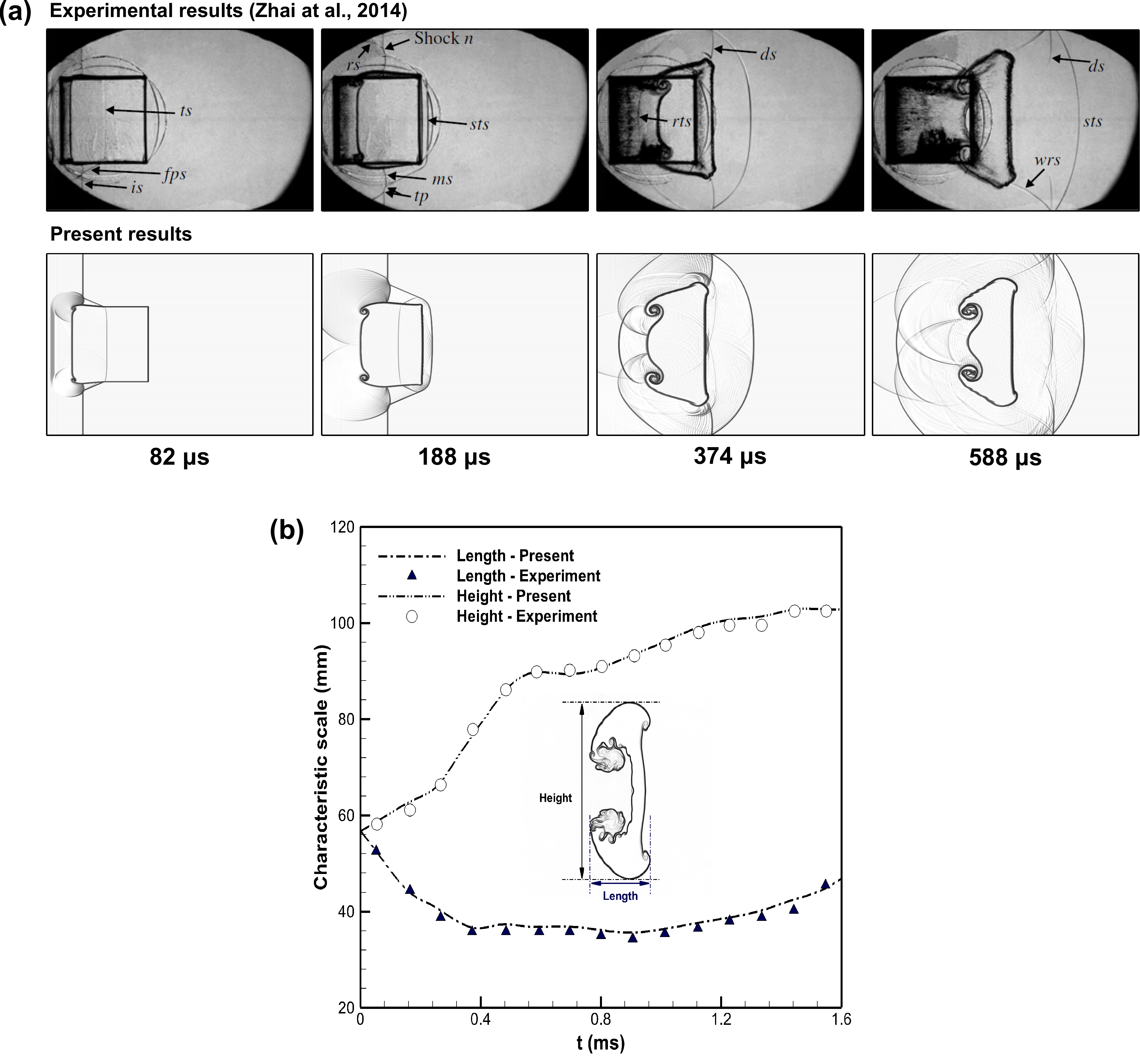}
	\caption{Validation of the numerical solver: (a) comparison of schlieren images between the experimental results of Zhai et al. \cite{zhai2014interaction} and the present numerical results at different time instants, and (b) the evolving interfaces between the experimental results and the present numerical results for a $\text{N}_{2}$ square bubble surrounded by $\text{SF}_{6}$ gas. The definitions of the interfacial characteristic scales are inserted.}
	\label{Fig:3}
\end{figure}

\subsection{Validation of the numerical solver}
\label{Sec:3.3}

A detailed validation study of the present numerical scheme for shock wave interaction with cylindrical and square bubbles was presented by Singh and co-authors \cite{singh2020role,singh2021nonequilibrium, singh2021bulk}, where good agreement was obtained for shock wave structures, positions, and bubble deformations. 
In the current study, the numerical results are validated with the experimental study of  Zhai et al. \cite{zhai2014interaction} for checking the validity of the present computational model, and the in-house developed explicit modal DG code, In this validation case, the gas square bubble is filled with nitrogen gas, while the ambient zone is composed of $\text{SF}_{6}$ gas. The benchmark simulation is computed on a weak planar shock wave of Mach number $\text{M}_{s}=1.28$. Figure \ref{Fig:3}(a) shows a comparison of the schlieren images between the experimental results of Zhai et al. \cite{zhai2014interaction} and the present numerical results at different times instants. These numerical simulations share the same initial condition, resolution, wave pattern, and diffusion layer thickness. The present schlieren images, including the vortex structures resembling one another, are in excellent agreement with the experimental results.
Furthermore, the time variations of the interfacial characteristic scales i.e., the length, and the height of the evolving interface for the $\text{N}_{2}$ square bubble, are also illustrated in Fig. \ref{Fig:3}(b). It can be seen from the plot that the present results, including the general trend of the interfacial characteristic scales changing with time, are found very close to the experimental results of Zhai et al. \cite{zhai2014interaction}.

\begin{figure} [hbt!]
	\centering
	\includegraphics[width=1.0\linewidth]{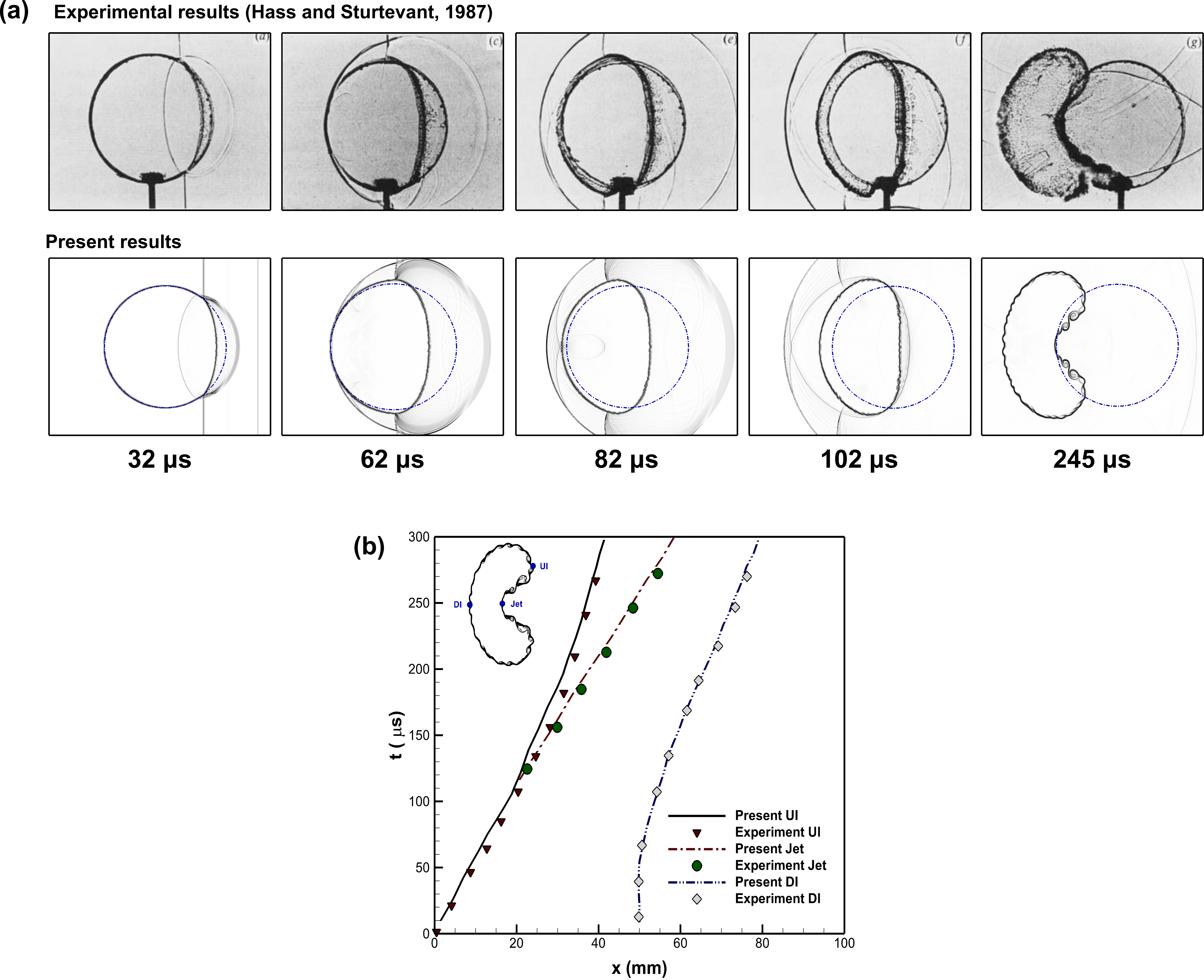}
	\caption{Validation of the numerical solver: comparison of (a)  numerical schlieren images between the experimental results of Haas and Sturtevant \cite{haas1987interaction} and the present numerical results at different time instants, and (b) the computed characteristic interface points (UI, DI, and Jet) between the experimental results and the present numerical results for a shock-accelerated $\text{R}_{22}$ cylindrical bubble surrounded by air. The definitions of the characteristic interface points are inserted.}
	\label{Fig:4} 
\end{figure}

Further, the current computational model is also validated through a comparison with the experimental results of Hass and Sturtevant\cite{haas1987interaction}, in which the cylindrical gas bubble was filled with refrigerant-22 $(\text{R}_{22})$ and the ambient zone was composed of air. The aforementioned experimental and computational studies also had a weak shock with $\text{M}_{s}=1.22$. Figure \ref{Fig:4}(a) compares the schlieren images between the experimental results \cite{haas1987interaction} and the present numerical results at different times. As seen from Fig. \ref{Fig:4}(a), the schlieren images are in good agreement across all experimental results. Furthermore, Figure \ref{Fig:4}(b) shows the space--time diagram for the characteristic interface points [i.e., upstream interface (UI), downstream interface (DI), and inward-jet head (jet)]. The numerical results are in good agreement with the experimental results \cite{haas1987interaction}. The positions and speeds of the various shock waves and the interfaces are accurately simulated by the computational model.

In Table \ref{tab:2}, the velocities of these characteristic interface points ($V_{DI}$, $V_{UI}$, and $V_{Jet}$), together with the time intervals involved in their computation, are presented alongside those obtained by Hass and Sturtevant \cite{haas1987interaction}. These velocities are estimated during the propagation inside the bubble and taken along the $x-$direction of the centerline of the domain. The discrepancy between the experimental, computational, and simulated velocities can be seen in Fig. \ref{Fig:4}(b), where the characteristic interface points do not exactly coincide. These discrepancies could be caused by the current numerical model ignoring the mass fraction term.

\begin{table}
	\caption{Validation of the numerical solver: comparison of velocities of characteristic interface points between experimental data, and the present numerical data. $V_{DI}$: velocity of downstream interface; $V_{UI}$: velocity of upstream interface; $V_{Jet}$: velocity of inward-jet head.}
		\label{tab:2} 
		\begin{ruledtabular}
			\begin{tabular}{cccc}
				Velocity & $V_{DI}$ (m/s) & $V_{UI}$ (m/s) & $V_{Jet}$ (m/s)  \\
				\hline
				Present	         & 148 & 175 &  228   \\
				Experimental	 & 145 & 170 &  230   \\
				Error (\%)	         & $-2.07$ & $-2.86$ & $0.87$  \\ 
			\end{tabular}
	\end{ruledtabular}
\end{table}

\section{Results and discussion: effect of Mach number on shock-accelerated square light bubble}
\label{Sec.4}

In this section, the effects of Mach numbers on the flow dynamics of the shock-accelerated square light bubble are investigated. The impacts of an initial interface perturbation on the flow morphology, wave patterns, vorticity distribution, interface movements, and qualitative analysis are emphasized. 
To investigate the Mach numbers effect on the shock-accelerated square light bubble, three different Mach numbers ($\text{M}_{s}=1.21, 1.7$, and 2.1) are selected for the numerical simulations. Helium is used as the bubble gas; this has been widely adopted as the light gas in studies of the RM instability. 

\begin{figure} [hbt!]
	\centering
	\includegraphics[width=1.0\linewidth]{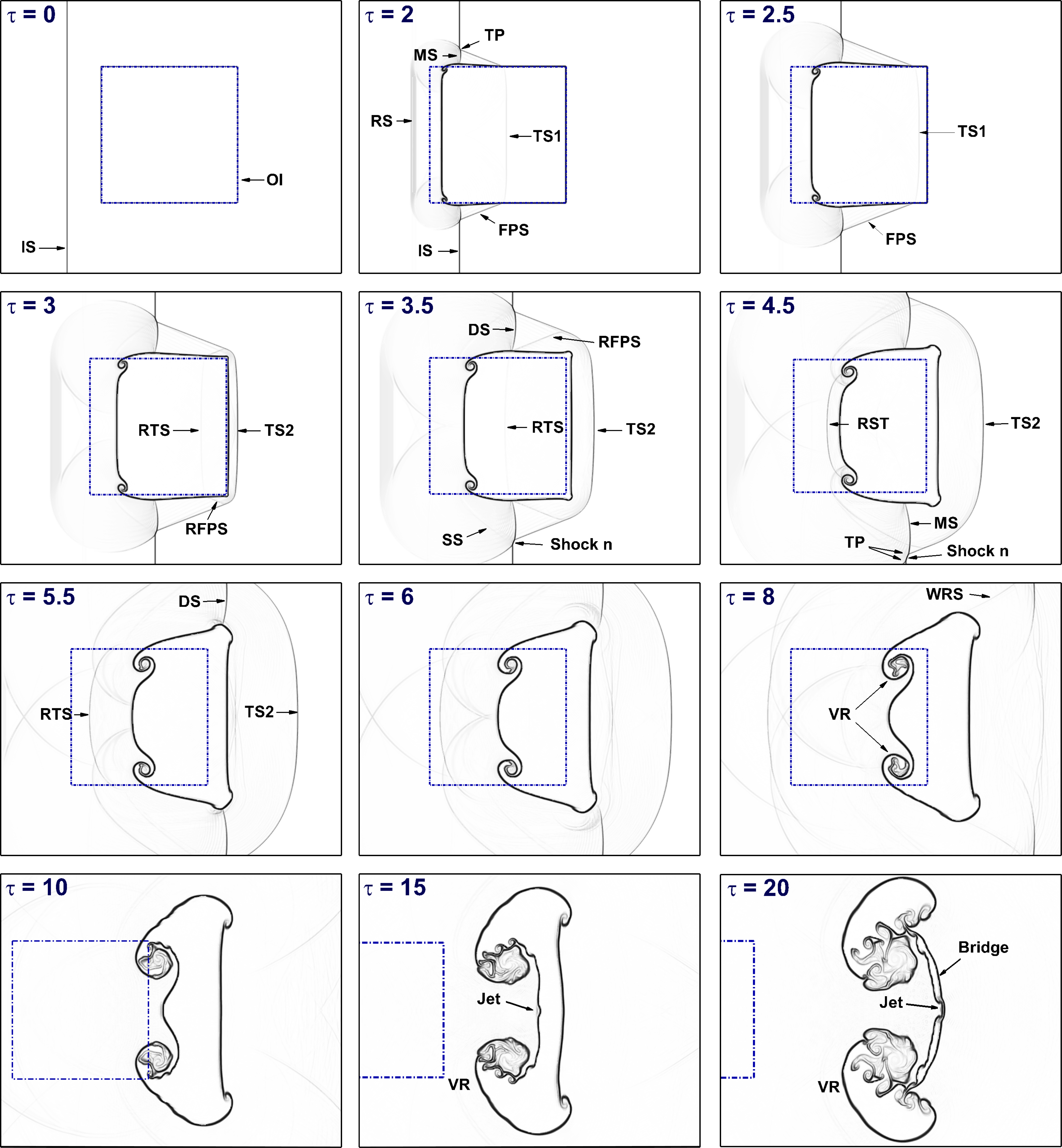}
	\caption{Overall visualization of the flow evolution in the shock-accelerated square helium bubble at $\text{M}_{s}=1.21$: evolution of numerical schlieren images at different times instants.}
	\label{Fig:5} 
\end{figure}

\subsection{Visualization of flow fields }
\label{Sec.4.1}

Figure \ref{Fig:5} shows an overall visualization of the flow evolution in the shock-accelerated square helium bubble with $\text{M}_{s}=1.21$ at different times instants. In this figure, a sequence of the numerical schlieren images of the square bubble accelerated by a planar incident shock (IS) wave (IS) is illustrated. Before interacting with the IS wave, the initial state of the bubble interface can be clearly observed $(\tau=0)$. When the IS wave travels along the upper and lower boundaries, the bubble starts to compress. Also, a transmitted shock (TS1) wave propagating downstream inside the bubble is generated, while, a reflected shock (RS) wave travels upstream simultaneously $(\tau=2)$. The propagation speed of the IS wave inside the bubble is smaller than that in the surrounding gas due to the small acoustic impedance. Therefore, the generated TS1 wave inside the bubble travels faster behind the IS wave. The TS1 wave is itself refracted at the bubble interface and transmits a new oblique shock wave so-called free precursor shock (FPS) in the ambient gas. The FPS and IS waves are then joined together and mutually modified producing a triple point (TP), a Mach stem (MS), and a shock outside the bubble. As the interaction develops, two small vortices are observed at the left corners of the square interface due to the vorticity deposition $(\tau=2.5)$. As the TS1 wave inside the bubble encounters the downstream interface, a secondary transmitted shock (TS2) wave traveling downward is produced and seen ahead of the original IS wave $(\tau=3)$.
Subsequently, a reflected transmitted shock (RTS) wave inside the square helium bubble is produced at the upstream surface and is moving now in the opposite direction, toward the bubble’s front. A new shock wave called reflected free precursor shock (RFPS) is also generated and connected between the RTS wave from the bubble rear surface and the FPS wave. This RFPS wave propagates toward the bubble frontal surface. A diffracted shock wave (DS) is also produced, moving along the rightmost surface $(\tau=3.5-6)$. Once the RTS wave within the bubble reaches the bubble front, it transmits into the surrounding gas in front of the bubble. The impact of the incident shock wave on the interface evolution is decreased as time goes on and the vortex pair (VR) at the corners gradually grows, caused by the produced vorticity $(\tau=8-10)$. Then, the evolving bubble interface starts to transform into a mushroom shape, and a re-entrant gas jet head is subsequently generated near the center of the bubble $(\tau=15)$. As time proceeds, the jet catches up with the downstream bubble interface, and then a pair of vortex rings (VR) connected with a bridge emerge and grow almost symmetrically $(\tau=15-20)$. At later times, the vortex pairs dominate the flow field entirely $(\tau=20)$.

\begin{figure} [hbt!]
	\centering
	\includegraphics[width=1.0\linewidth]{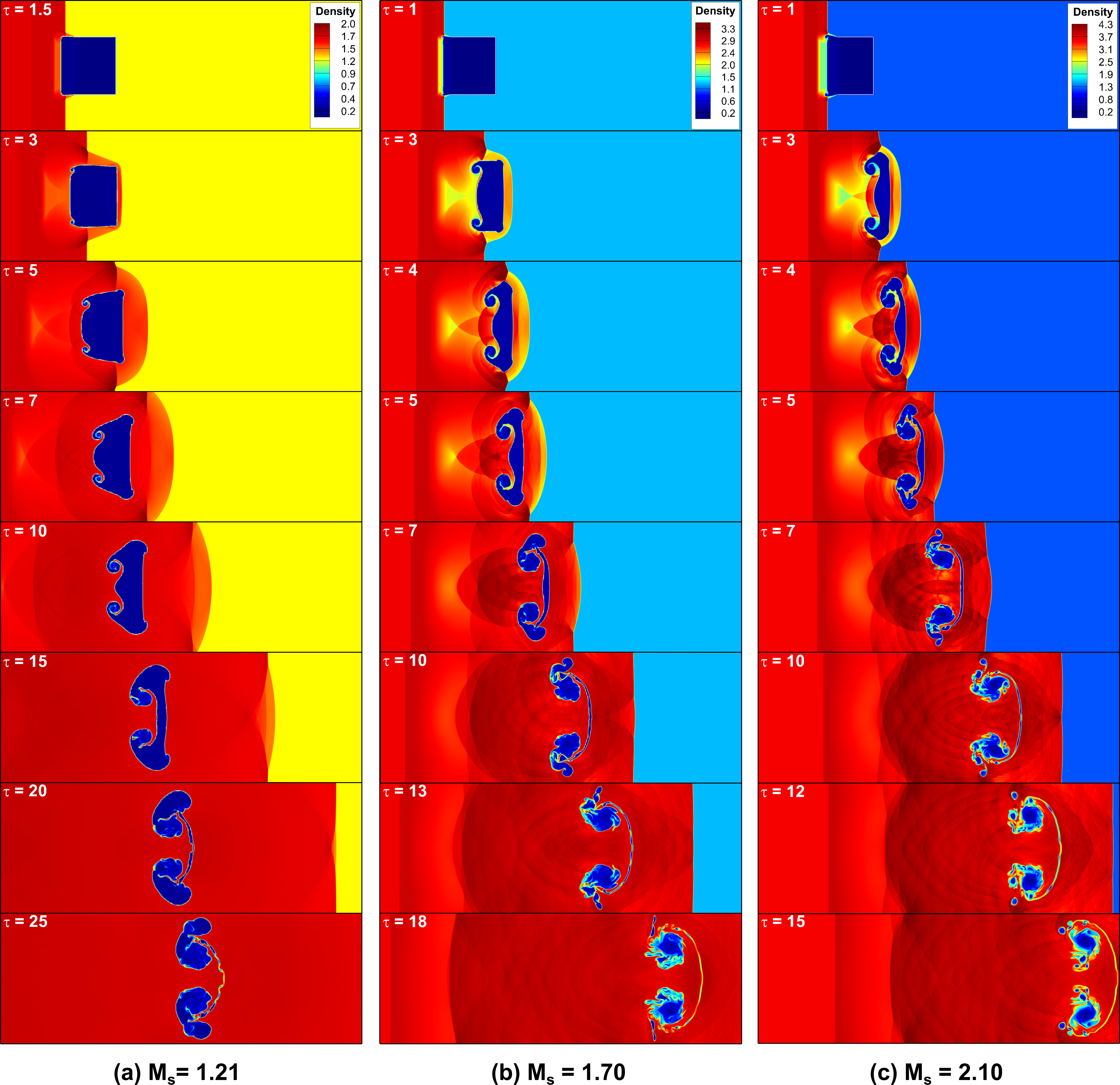}
	\caption{Effect of Mach number on the shock-accelerated square helium bubble: contours of density distribution for (a) $\text{M}_{s}=1.21$, (b) $\text{M}_{s}=1.7$, and (c) $\text{M}_{s}=2.1$ at different times instants.}
	\label{Fig:6} 
\end{figure}

Figure \ref{Fig:6} illustrates the effects of three different Mach numbers $(\text{M}_{s}=1.21,1.7, 2.1)$ on the time-dependent flow morphology of the shock-accelerated square helium bubble through density contours. For the three cases, due to the small acoustic impedance in the helium bubble, the TS1 wave inside the bubble moves faster than the outside IS wave. Similar to $\text{M}_{s}=1.21$, at higher Mach numbers, an irregular reflection process, including FPS wave, TP, and MS are formed at early stages. After hitting the downstream interface by TS1 wave, the RTS, RFPS, DS, and TS2 waves are also generated in all three cases. Subsequently, a re-entrant gas jet head is found in the centerline near the right interface of the bubble. It can be observed that the high shock Mach number causes a stronger interaction between the shock and bubble. The height of generated jet increases with the increasing the strength of the incident shock wave due to the larger expansion of the bubble interface upwards, as seen in Figs. \ref{Fig:6}(b)-(c). Furthermore, as the Mach number increases, the bubble deforms significantly. As a result, the generated wave patterns become more complex, and the size of the bubbles decreases noticeably. The re-entrant jet structure at $\text{M}_{s}=2.1$ is observed as the longest among three Mach numbers. Additionally, the size and strength of the rolled-up vortices increase significantly at high Mach numbers, and these vortices are conspicuous at the interface between the bubble and surrounding gas due to the baroclinic vorticity deposition.

\begin{figure} [hbt!]
	\centering
	\includegraphics[width=1.0\linewidth]{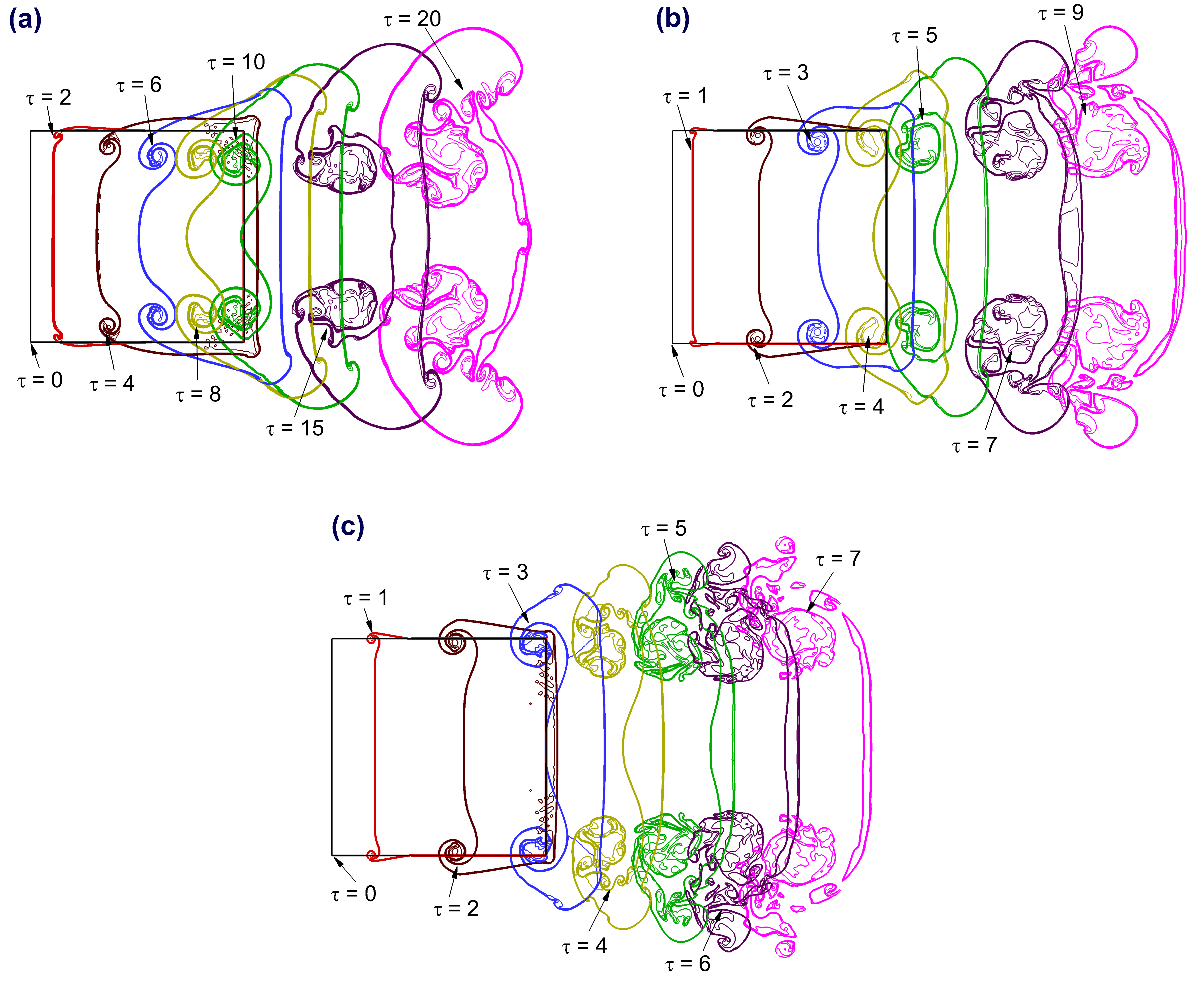}
	\caption{Effect of Mach number on the shock-accelerated square helium bubble: evolution of the bubble shape showing early compression for (a) $\text{M}_{s}=1.21$, (b) $\text{M}_{s}=1.7$, and (c) $\text{M}_{s}=2.1$.}
	\label{Fig:7} 
\end{figure}

Further, the evolution of the bubble shape during the interaction with the IS wave, shown in Fig. \ref{Fig:7}, illustrates the effect of Mach number on the shock-accelerated square helium bubble. The bubble appears to be compressed by the IS wave along the $x-$direction, and its top and bottom edges have been pushed forward near the horizontal axis of symmetry as compared to its middle section. This compression begins as soon as the IS wave reaches the upstream end of the bubble at the start of the interaction. The bubble appears to be compressed by the IS wave along the $x-$direction, and, compared with its middle section, the top and bottom edges have been pushed forward near the horizontal axis of symmetry. At the beginning of the interaction, this compression starts as soon as the IS wave hits the upstream end of the bubble. For all three cases, both upstream and downstream interfaces travel fast, and the upstream side presses inward under the influence of the IS wave at the early instants, as shown in Figs. \ref{Fig:7}(a)-(c). After the interaction, a pair of small vortices are generated at the left corners of the square interface due to the vorticity deposition. The size of these two vortices continuously grows over time. Besides it, the upper and lower horizontal interfaces of the bubble fold inward toward the upstream axis, and the bubble deforms into a divergent shape. Some small-scale rolled-up vortices are also generated on the upper and lower interfaces due to baroclinic vorticity generation. In the case of $\text{M}_{s}=1.21$, the size of these rolled-up vortices on the bubble interface is smaller in comparison to higher Mach numbers. It may be observed that as time goes, the vortex pair at the corners gradually grows. At later stages, the flow field is completely controlled by the vortex pairs.

\begin{figure} [hbt!]
	\centering
	\includegraphics[width=1.0\linewidth]{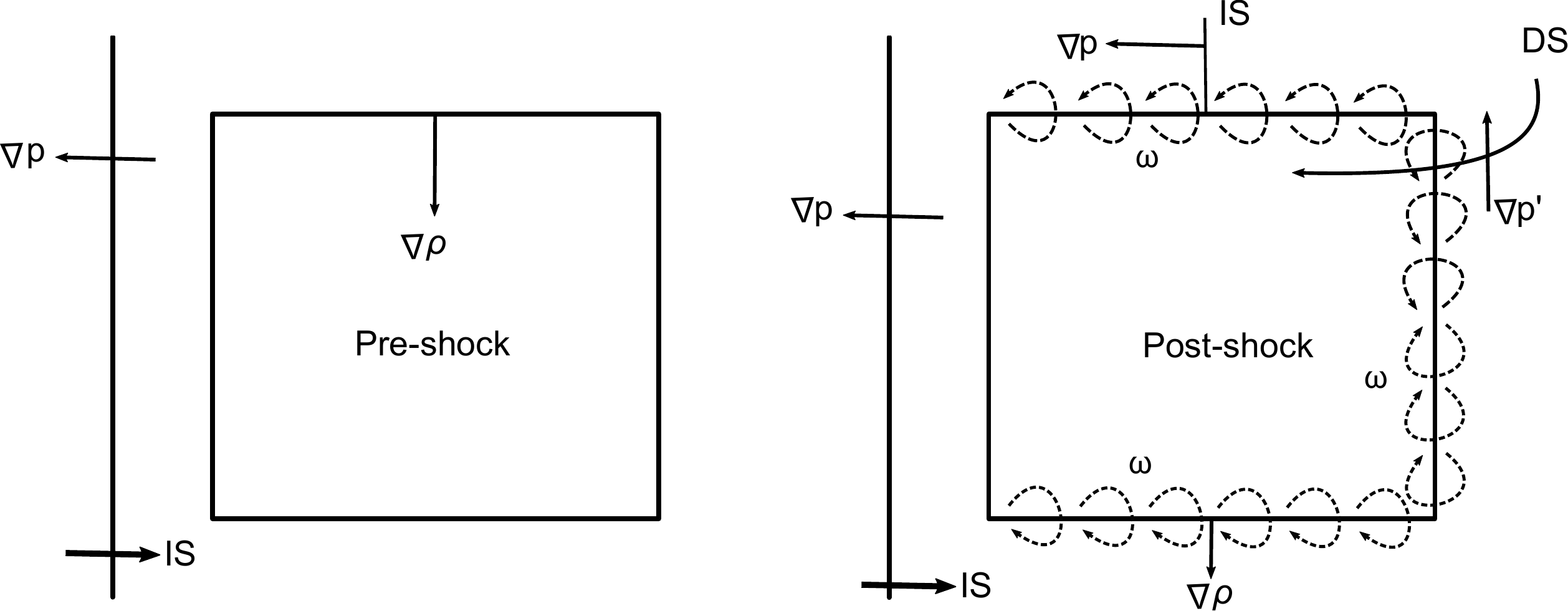}
	\caption{Schematic diagram of the vorticity generation on the interface of the square light bubble during and after the initial shock wave transits.}
	\label{Fig:8} 
\end{figure}

\subsection{Dynamics of vorticity production}
\label{Sec.4.2}

Vorticity production is an important mechanism in transport phenomena like turbulent mixing and noise generation. In shock-bubble interaction, the vorticity is generated in the flow fields and distributed initially on the bubble interface due to the misalignment of the pressure and density gradients when the IS wave passes. The vorticity transport equation explains the fundamental physics that occurred during an interaction process. 
Recently, a completely new viscous compressible vorticity transport equation, which includes several physically distinctive quantities has been derived by Myong \cite{singh2021bulk}. This equation can be rewritten without bulk viscosity $(\text{i.e. }f_{b}=0)$ as

\begin{equation}
	\label{Eq:13}
	\frac{D {\omega}}{D \tau} = \underbrace{({\omega} \cdot \nabla) \mathbf{u}}_{\text{P}_{\omega,str}}
	- \underbrace{{\omega}(\nabla \cdot \mathbf{u})}_{\text{P}_{\omega,dil}}
	+ \underbrace{\frac{1}{\rho^{2}} (\nabla \rho \times \nabla p)}_{\text{P}_{\omega,bar}}
	+ \underbrace{\frac{\mu}{\rho} \left[\nabla^{2} {\omega} 
		-\frac{1}{\rho}(\nabla \rho \times \nabla^{2}\mathbf{u})- \frac{1}{3} \nabla \rho \times \nabla (\nabla \cdot \mathbf{u})\right]}_{\text{P}_{\omega,vis}},
\end{equation}
where $\mathbf{u}$ is the velocity, ${\omega}=\nabla \times \mathbf{u}$ is the vorticity, $\rho$ is the density, $p$ is the pressure, and $\mu$ is the shear viscosity.
On the right-hand side of Eq. (\ref{Eq:13}), the term $({\omega} \cdot \nabla) \mathbf{u}$ 
represents the stretching or tilting of vorticity due to the flow velocity gradients, 
which is critical for three-dimensional turbulence and mixing. The term $ {\omega}(\nabla \cdot \mathbf{u})$ expresses the stretching of vorticity due to flow compressibility.
The term $(1/\rho^{2}) (\nabla \rho \times \nabla p)$ denotes the baroclinic vorticity production term, which is responsible for the generation of small-scale rolled-up vortices at the bubble interface. Moreover, this term is most prominent at the top and bottom ends of a vertical bubble due to the extreme misalignment of the density and pressure gradients.
The term $(\mu/\rho) \nabla^{2} {\omega}$ represents the rate of change of $\omega$ due to molecular diffusion of vorticity. The term $(\mu/ \rho^{2})(\nabla \rho \times \nabla^{2}\mathbf{u})$ represents the vorticity generated by the combination of density and velocity diffusion gradients. Finally, the last term $(\mu/ \rho^{2})(\nabla \rho \times \nabla (\nabla \cdot \mathbf{u}))/3$ represents the vorticity generated by the combination of density and viscous normal stress gradients, which resemble the baroclinic vorticity production generated by the combination of density and pressure gradients. 

\begin{figure} [hbt!]
	\centering
	\includegraphics[width=1.0\linewidth]{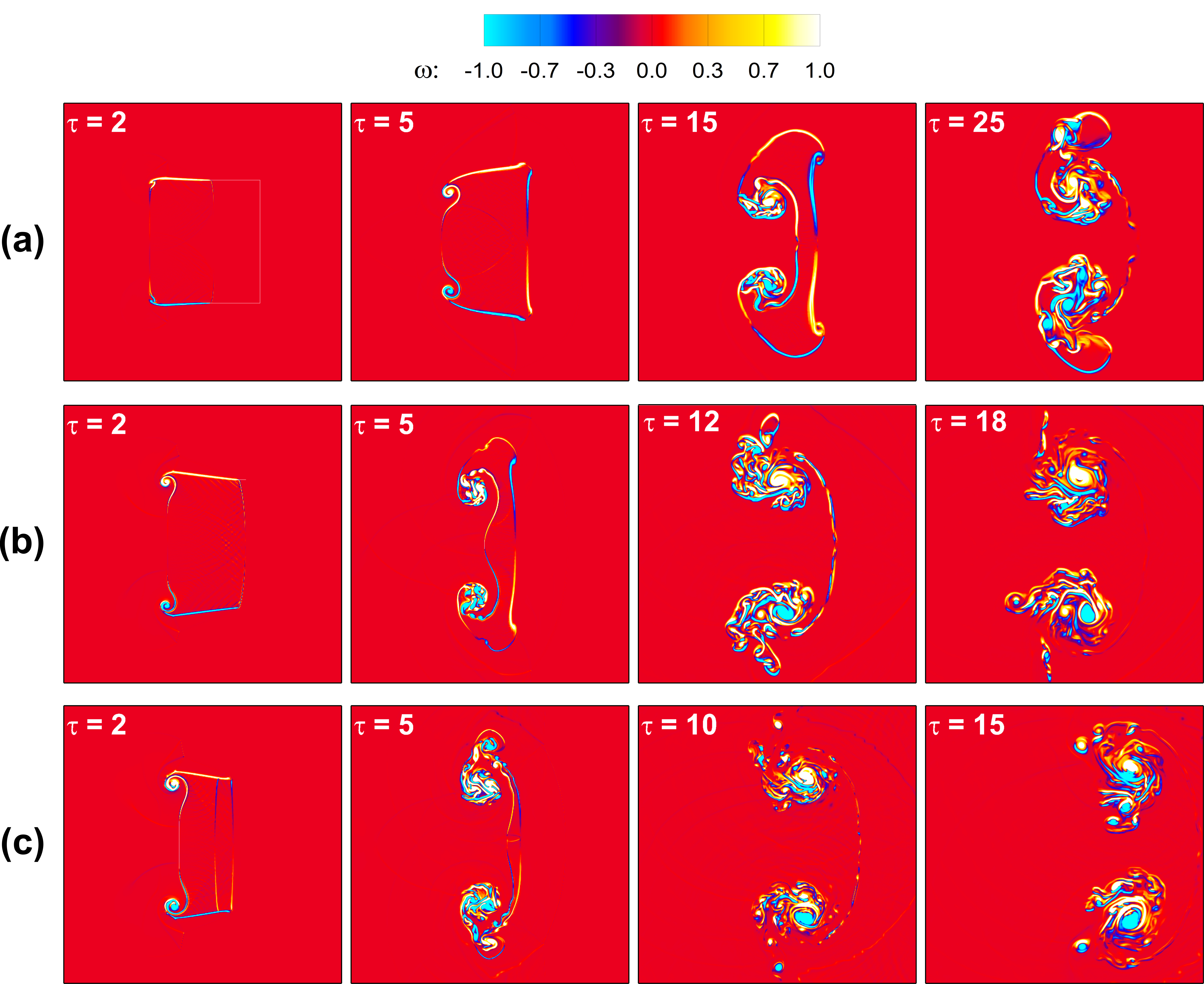}
	\caption{Effect of Mach number on the shock-accelerated square helium bubble: contours of vorticity distribution for (a) $\text{M}_{s}=1.21$, (b) $\text{M}_{s}=1.7$, and (c) $\text{M}_{s}=2.1$ at different times instants.}
	\label{Fig:9} 
\end{figure}

The deposition of baroclinic vorticity on the bubble interface is well recognized as a key factor causing the initial disturbance to grow. Now we insight how the baroclinic vorticity term affects the IS and TS1 waves as they pass through the stationary bubble interface early in their evolution. In the shock-accelerated bubble, the dominant pressure gradient occurs in the plane IS wave, while the dominant density gradient can be found at the bubble interface. When the plane IS wave passes over the bubble, it does not deform the bubble significantly. At the leftmost vertical interface, where the pressure and density gradients are perfectly aligned, a small quantity of vorticity is generated at the corners (top and bottom) when the IS wave touches them. A schematic diagram of the vorticity generation on the square bubble interface after the initial IS wave transits across the light gas bubble is illustrated in Fig. \ref{Fig:8}. As the incident shock propagates along with the horizontal upper interface, Mach reflection occurs, in which the Mach stem (MS) connects the IS wave with the square interface. Therefore, the MS contributes the pressure gradient for the vorticity generation on the interface, and the baroclinic vorticity term is thus triggered gradually as the IS wave travels upwards over the square interfaces. 

Figure \ref{Fig:9} illustrates the effect of Mach number on the vorticity distribution of the shock-accelerated square helium bubble at different times instants.
Initially, the vorticity is equal to zero everywhere. When the IS wave passes across the bubble, the baroclinic vorticity is mainly deposited locally on the bubble interface in the early stage, where the discontinuity between helium gas and the ambient gas exists. 
At the top and bottom locations of the bubble where the density and pressure gradients are orthogonal, the magnitude of the vorticity is maximum and it is zero at the interface along the axis of the bubble where the density and pressure gradients are collinear. A significant quantity of positive vorticity is generated on the upper horizontal side of the bubble interface, while a significant quantity of  negative vorticity is generated on the lower horizontal side of the bubble, as shown in Figs. \ref{Fig:9}(a)-(c). This is because of the IS wave propagating from left to right along with the bubble interface. As a result, the density gradient is everywhere radially outwards at the bubble interface and the pressure gradient is across the upstream IS wave. Furthermore, the vortical structure in the upper interface is observed with positive vorticity in the center, surrounded by tails of negative vorticity, while vise-versa situations are noticed in the bottom interface of the square bubble. One can observe that there are significant differences in vorticity distribution for the different Mach numbers after the interaction. For $\text{M}_{s}=1.21$, a small quantity of vorticity is generated around the rolled-up vortices on the bubble interface, as shown in Fig. \ref{Fig:9}(a). These rolled-up vortices are more pronounced for high Mach numbers, as seen in Figs. \ref{Fig:9}(b)-(c). In summary, the generation and distribution of vorticity play a dominant role at high Mach numbers when rolled-up vortices are formed.

\begin{figure} [hbt!]
	\centering
	\includegraphics[width=1.0\linewidth]{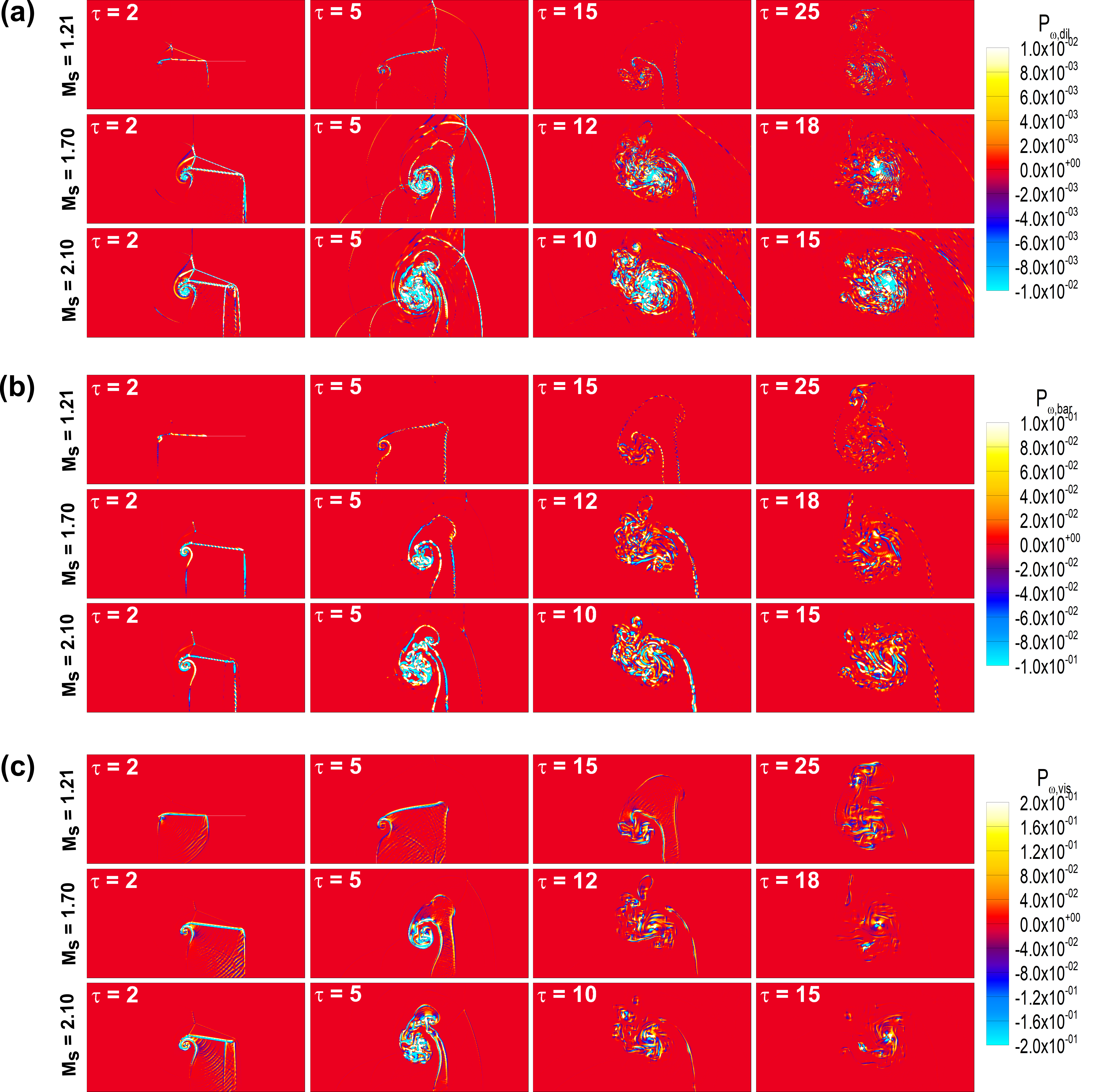}
	\caption{Effect of Mach number on the shock-accelerated square helium bubble: contours of (a) dilatational vorticity production $(\text{P}_{\omega, dil})$, (b) baroclinic vorticity production $(\text{P}_{\omega, bar})$, and (c) viscous vorticity production $(\text{P}_{\omega, vis})$ terms.}
	\label{Fig:10} 
\end{figure}

Figure \ref{Fig:10} illustrates the effect of Mach number on different expressions appearing in the right-hand side of Eq. (\ref{Eq:13}) at different time intervals. The results show that the all vorticity production terms obtain significant values during the interaction and diminish shortly after passing the shock wave. As one can see, the viscous production term $(\text{P}_{\omega, vis})$ has the most dominant mechanism, followed by baroclinic vorticity and dilatational vorticity production terms in the interaction for strong shock waves. Interestingly, It can  be also seen from the dilatational production term $(\text{P}_{\omega, dil})$ contours that there are locally stretched structures in the core region of the vortex due to the compressibility effect originating from local regions of compression and expansion. Moreover, the existence of evolving large scale vortices which interact with the different shock patterns present in the flow and finally split into small-scale vortices.

The vorticity at the bubble interface is critical for gas mixing inside and outside the bubble. Therefore, four important spatially integrated fields are investigated in detail to obtain a better understanding for Mach number effects on the physics of vorticity production: (i) average vorticity $(\omega_{av})$, (ii) dilatational vorticity production $(P_{\omega, {dil}})$, (iii) baroclinic vorticity production $(P_{\omega, {bar}})$, and (iv) viscous vorticity production $(P_{\omega, {vis}})$. The spatially integrated field of average vorticity is defined as
\begin{equation}
	\label{Eq:14}
	\omega_{av} (\tau) = \frac{\int_{D} |\omega| dx dy}{\int_{D} dx dy},
\end{equation}
where $D$ represents the entire computational domain. The spatially integrated field of dilatational vorticity production term is computed as follows:
\begin{equation}
	\label{Eq:15}
	P_{\omega, {dil}}(\tau) = -\frac{\int_{D} \left | {\omega}(\nabla \cdot \mathbf{u}) \right| dx dy}{\int_{D} dx dy}.
\end{equation}
The spatially integrated field of baroclinic vorticity production is given by
\begin{equation}
	\label{Eq:16}
	P_{\omega, {bar}}(\tau) = \frac{\int_{D} \left | \frac{1}{\rho^{2}} (\nabla \rho \times \nabla p) \right| dx dy}{\int_{D} dx dy}.
\end{equation}
Finally, the spatially integrated field of viscous vorticity production is defined as
\begin{equation} 
	\label{Eq:17}
	P_{\omega, {vis}}(\tau) = \frac{\int_{D} \left | \frac{\mu}{\rho} \nabla^{2} {\omega}
		-\frac{\mu}{\rho^{2}} (\nabla \rho \times \nabla^{2}\mathbf{u})- \frac{1}{3}   \frac{\mu}{\rho^{2}} \nabla \rho \times \nabla (\nabla \cdot \mathbf{u}) \right| dx dy}{\int_{D} dx dy}.  
\end{equation}

\begin{figure} [hbt!]
	\centering
	\includegraphics[width=1.0\linewidth]{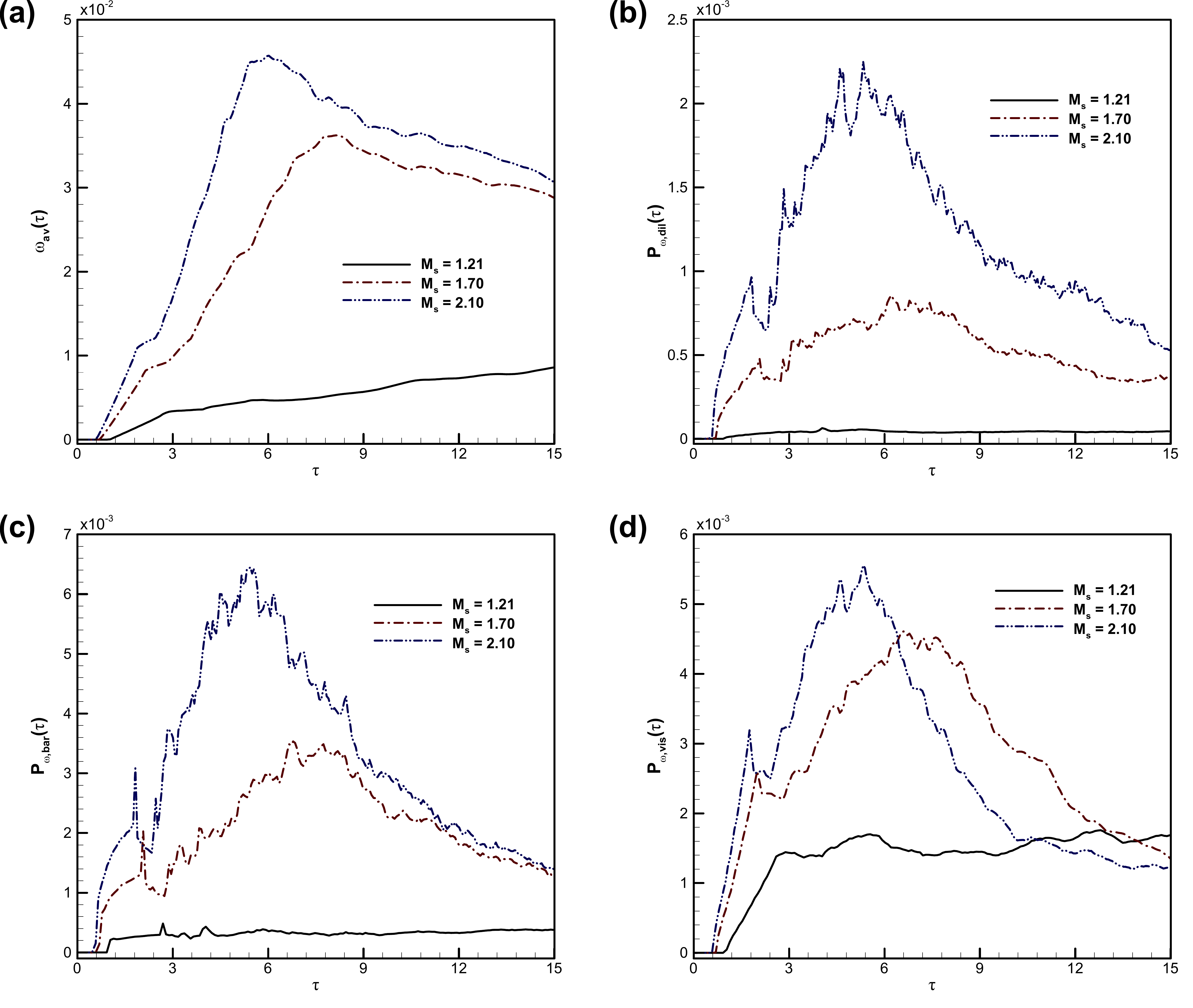}
	\caption{Effect of Mach number on the shock-accelerated square helium bubble: spatially integrated fields of (a) average vorticity $(\omega_{av})$, (b) dilatational vorticity production $(P_{\omega, {dil}})$, (c) baroclinic vorticity production $(P_{\omega, {bar}})$, and (d) viscous vorticity production $(P_{\omega, {vis}})$.}
	\label{Fig:11} 
\end{figure}

Figure \ref{Fig:11} illustrates the effect of Mach number on the spatially integrated fields of average vorticity, absolute dilatational vorticity, absolute baroclinic vorticity, and absolute viscous vorticity in the shock-accelerated square helium bubble. It can be observed from Fig. \ref{Fig:11}, these spatially integrated fields in case of $\text{M}_{s}=1.21$ is the smallest among the three cases when the incident and reflected shock waves collide with the bubble. The spatially integrated fields are substantially enhanced in the case of $\text{M}_{s}=2.1$. For all three Mach numbers, the spatially integrated fields increase with time, which implies that the ambient gas is increasingly entrained into the distorted square helium bubble. The vortices produced by the shock wave--bubble interaction encourage the mixing of ambient gas with the square helium bubble. When the reflected shock waves impinge on the distorted helium bubble again, the spatially integrated fields exhibit their greatest growth rate, which indicates that the vorticities are significantly enhanced during this period, as shown in Fig. \ref{Fig:11}. The growth rate then slows under the influence of the higher viscosity in the flow field. As a result, the evolution of the spatially integrated fields for vorticity and its associated components does not exhibit a simple monotonic relationship with the ambient gas.

\subsection{Evolution of enstrophy and dissipation rate}
\label{Sec.4.3}

By evaluating the time evolution of the enstrophy, the physical phenomenon of vorticity generation during the interaction process can be better described. The time evolution of the enstrophy can be defined as the spatial integral of the square of the vorticity in the flow field:
\begin{equation}
	\label{Eq:18}
	\Omega (\tau) = \frac{1}{2} \int_{D} \omega^{2} dx dy.
\end{equation}
Moreover, the Mach number effects can be investigated by introducing the area-weighted dissipation rate of kinetic energy:
\begin{equation}
	\label{Eq:19}
	\epsilon (\tau) = - \int_{D} (\Pi_{xx} S_{xx} + \Pi_{xy} S_{xy} + \Pi_{yx} S_{yx} +  \Pi_{yy} S_{yy}) dx dy,
\end{equation}
where $\Pi_{ii}$ is the viscous shear stress, and $S_{ij}$ is the strain rate, defined as $S_{ij}= \frac{1}{2}(\partial u_{i}/\partial x_{j} + \partial u_{j}/\partial x_{i})$.

\begin{figure} [hbt!]
	\centering
	\includegraphics[width=1.0\linewidth]{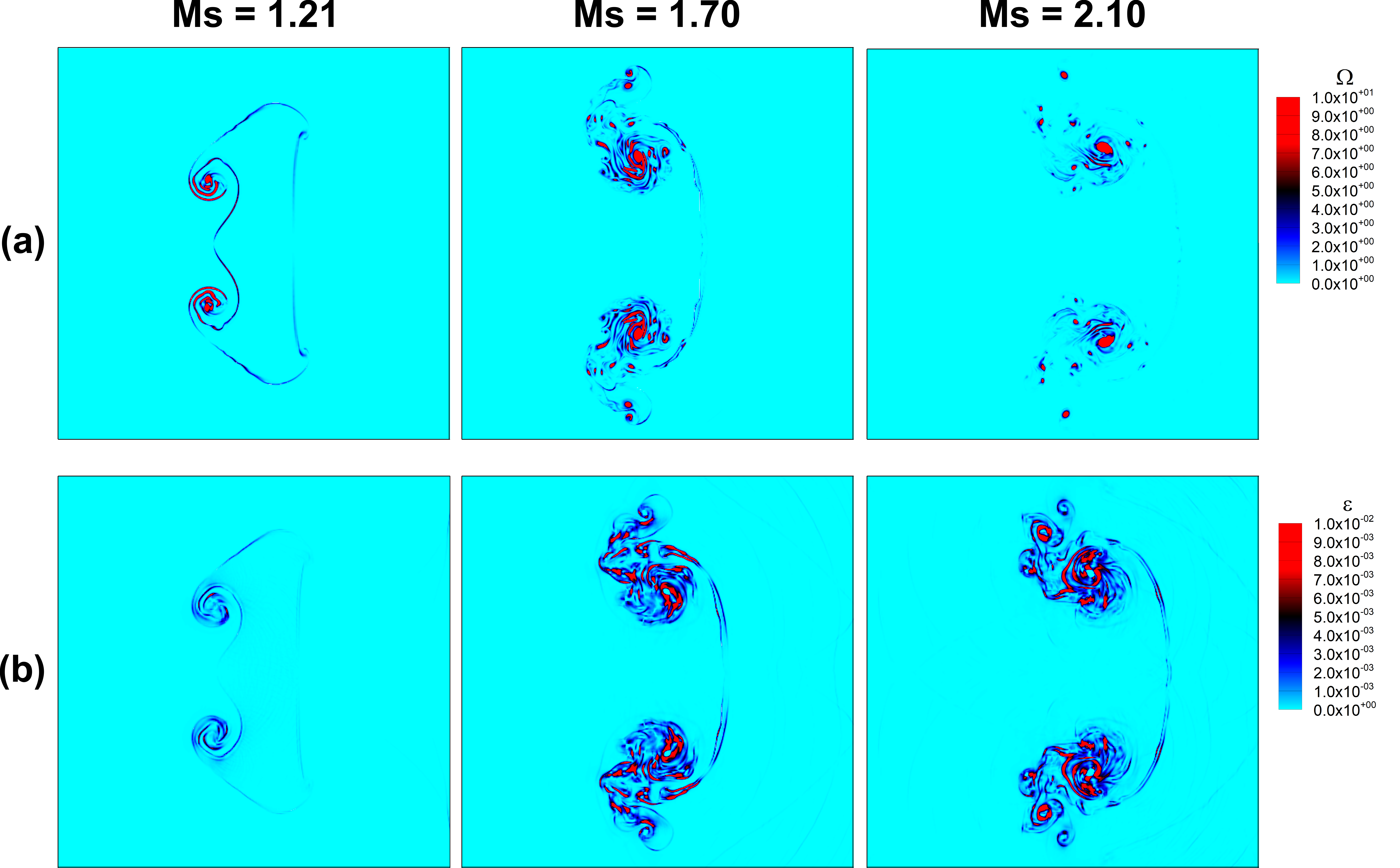}
	\caption{Effect of Mach number on the shock-accelerated square helium bubble: contours of (a) enstrophy, and (d) dissipation rate at time $\tau =10$.}
	\label{Fig:12} 
\end{figure}
\begin{figure} [hbt!]
	\centering
	\includegraphics[width=1.0\linewidth]{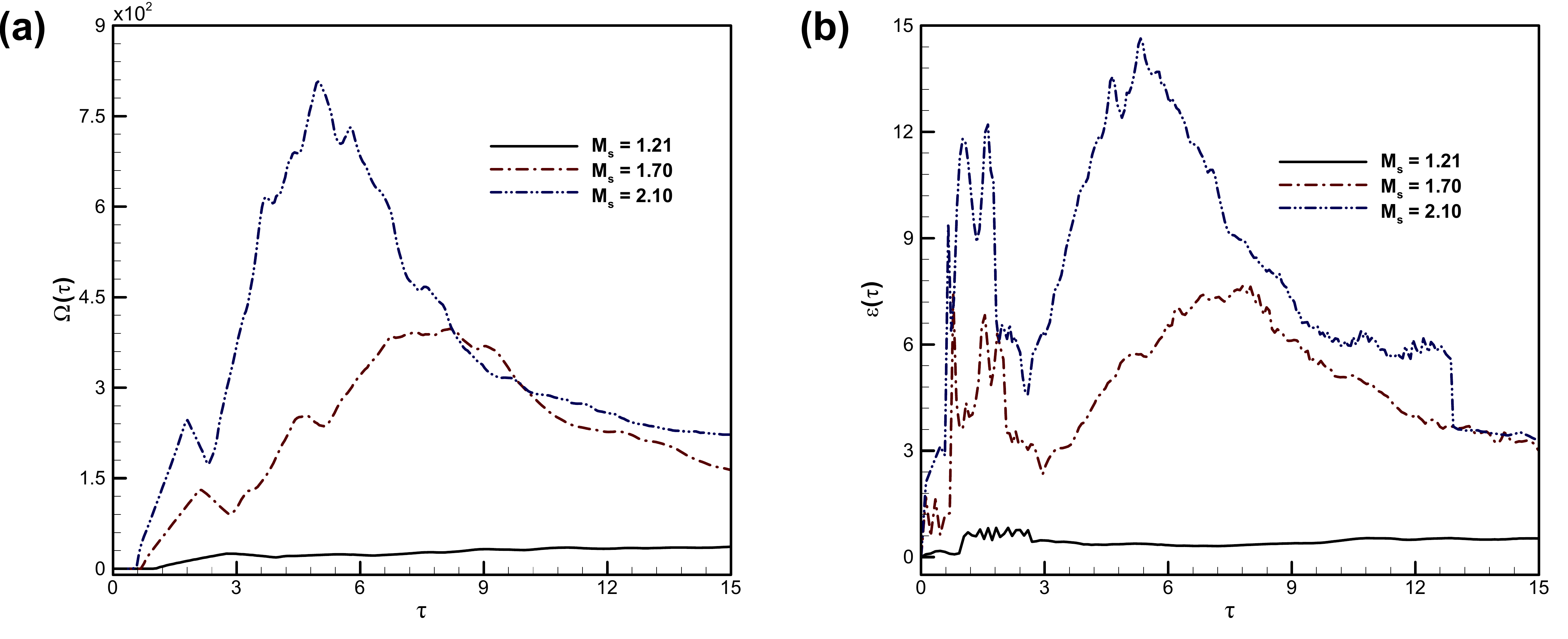}
	\caption{Effect of Mach number on the shock-accelerated square helium bubble: spatially integrated fields of (a) enstrophy $(\Omega)$, and (b) dissipation rate $(\epsilon)$ during interaction process.}
	\label{Fig:13} 
\end{figure}

Figure \ref{Fig:12} outlines the effect of Mach number on the enstrophy and dissipation rate in the shock-accelerated square helium bubble at a time instant $\tau =10$. After the interaction, there are considerable differences in enstrophy and dissipation rate  at various Mach numbers. A significant amount of the enstrophy and dissipation rate can be observed inside the rolled-up vortices of the deformed bubble interface. At high Mach numbers, enstrophy and dissipation rate are substantially enhanced compared to low Mach number. The spatially integrated fields of the enstrophy and dissipation rate over time are illustrated in Fig. \ref{Fig:13} to further investigate the Mach number effects on the vorticity production and kinetic energy. The enstrophy is zero until the shock wave reaches the upstream pole of the bubble. Baroclinic vorticity production leads to an increase during the shock wave passage. A first local maximum in enstrophy is reached after the shock has passed half of the bubble, an effect that can be observed for all simulations. 
Thereafter, a slight decay is visible, followed by another increase due to shock transmission and shock reflections at the interface. Subsequently, the enhanced vorticity promotes the mixing of gases inside and outside the gas bubble, and thus accelerates the transfer and consumption of vorticity energy, which gradually weakens the enstrophy intensity in the bubble region, as shown in Fig. \ref{Fig:13}(a). The same phenomena is observed in all shock Mach numbers. Only overall enstrophy levels differ as stronger shock waves generate more enstrophy. This similar physcis can also be seen in the time evolution of the dissipation rate, as shown in Fig. \ref{Fig:13}(b). At low Mach number, the dissipation rate remains relatively constant over time during the entire weak interaction process, whereas the dissipation rate at high Mach numbers experiences a substantial increase during the strong interaction process. Furthermore, during the interaction process, a non-monotonic pattern in enstrophy and dissipation rate emerges.

\begin{figure} [hbt!]
	\centering
	\includegraphics[width=1.0\linewidth]{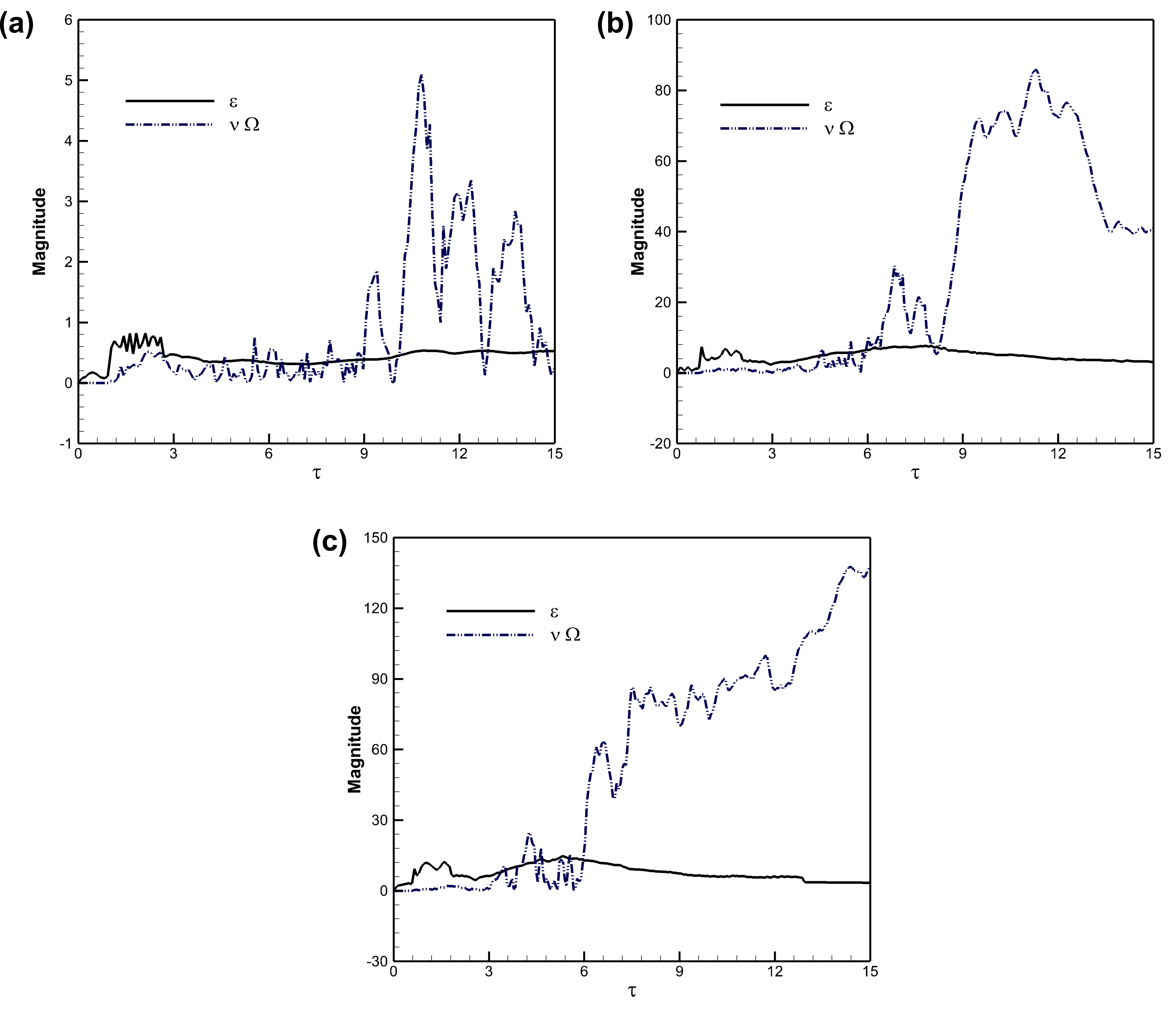}
	\caption{Comparison between the magnitudes of spatially integrated fields of $\epsilon$ and $\nu \Omega$ during interaction process at $\text{M}_{s}=1.21$, (b) $\text{M}_{s}=1.7$, and (c) $\text{M}_{s}=2.1$.}
	\label{Fig:14} 
\end{figure}

In homogeneous flows, the kinetic energy $(\epsilon)$ is directly proportional to the enstrophy $(\Omega)$ i.e. $\epsilon \approx \nu \Omega$, where $\nu$ is the kinematic viscosity \cite{foias2001navier}. In shock-accelerated interface problems, the flow fields are typically inhomogeneous. In such problems, inhomogeneous flows should be expected to differ intrinsically from homogeneous flows. To verify this phenomena, a comparison between the magnitudes of spatially integrated fields of $\epsilon$ and $\nu \Omega$ during interaction process at three different Mach numbers is conducted, as shown in Fig. \ref{Fig:14}. The numerical results show that the kinetic energy declines proportional to the kinematic viscosity $\nu$  times the enstrophy. Surprisingly, when the strength of the shock wave increases, the disparity between the magnitudes of these numbers widens dramatically, which confirms that the present flow configuration during the interaction between the IS wave and square light bubble is inhomogeneous.

\subsection{Shock trajectories and interface features}
\label{Sec.4.4}

\begin{figure} [hbt!]
	\centering
	\includegraphics[width=1.0\linewidth]{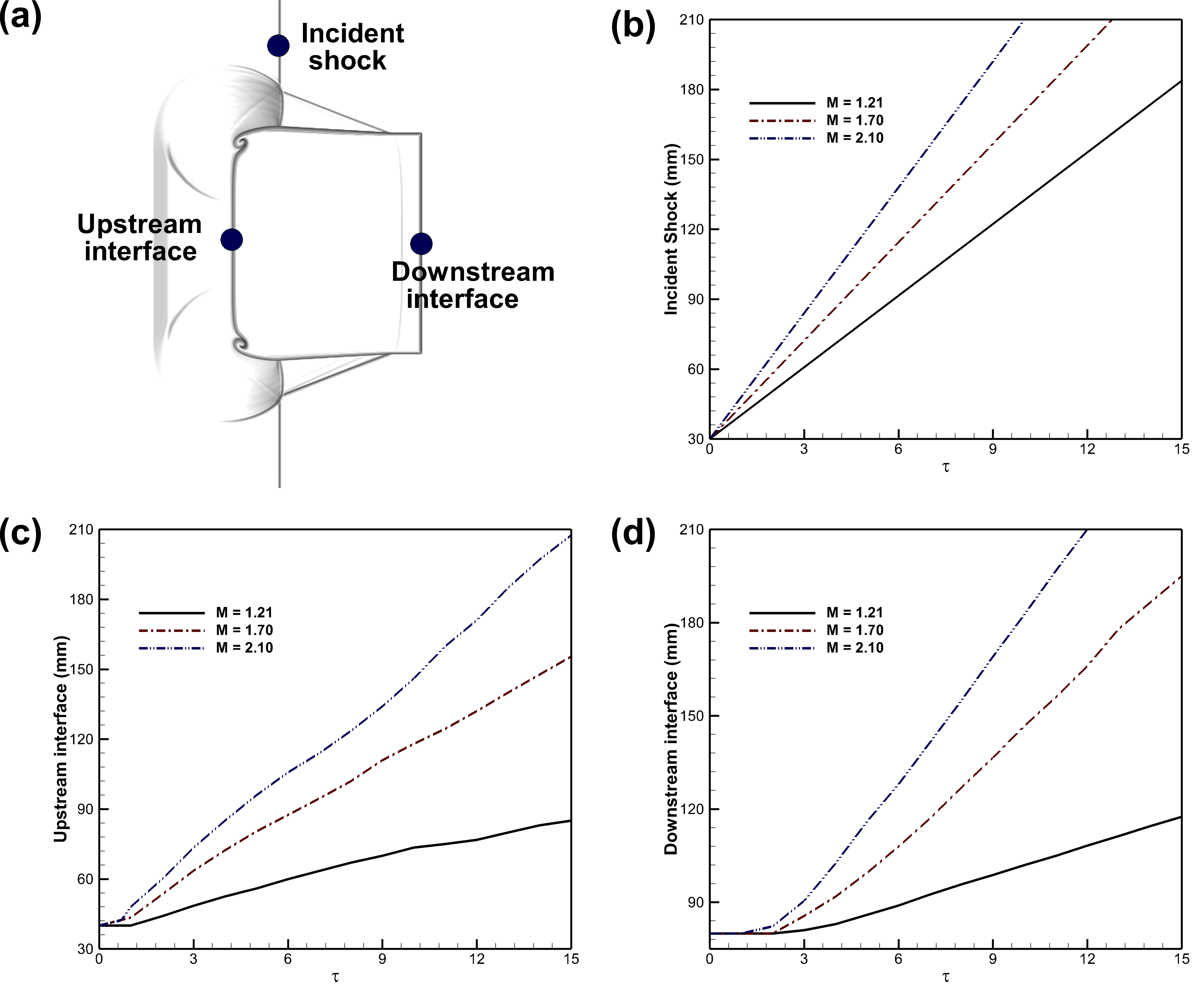}
	\caption{Effect of Mach number on the shock-accelerated square helium bubble: (a) 
		schematic diagram with the shock trajectory points, (b) incident shock wave based trajectories, (b) upstream interface based trajectories, and (c) downstream interface based trajectories.}
	\label{Fig:15} 
\end{figure}

Further, a quantitative analysis based on the physical phenomena of the shock trajectories, and the interface features are presented here to investigate the effect of Mach number on the shock-accelerated square helium bubble. Figure \ref{Fig:15}(a) shows a schematic diagram of the shock trajectory points (incident shock; upstream interface; and downstream interface) on the square bubble at the middle stage of the interaction process. Figures \ref{Fig:15}(c)-(d) illustrate the effect of Mach number on the shock trajectory points indicated in Fig. \ref{Fig:15}(a). It can be observed from the figure that the fastest displacement of these shock trajectory points occurs at high Mach number $\text{M}_{s}=2.1$, while the slowest displacement is found at low Mach number $\text{M}_{s}=1.21$.

\begin{figure} [hbt!]
	\centering
	\includegraphics[width=1.0\linewidth]{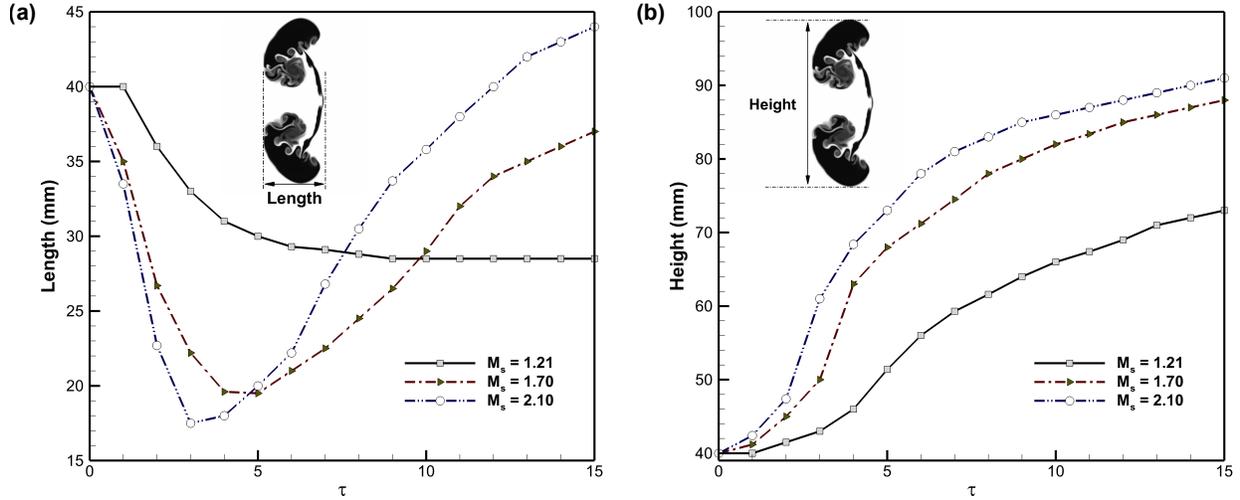}
	\caption{Effect of Mach number on the shock-accelerated square helium bubble: temporal variations of the interfacial characteristic scales$-$(a) the length, and (b) the height$-$ of the evolving interface for the computed bubble. The definitions of the length and the height of the evolving interface are inserted.}
	\label{Fig:16} 
\end{figure}

Figure \ref{Fig:16} illustrates the effect of Mach number on the temporal variations of the interfacial characteristic scales (i.e., the length and height of the evolving interface) for the computed square helium bubble. The length and height of the evolving interface are defined in the figure. Early shock compression rapidly reduces the length of the evolving interface after the incident shock arrives for all Mach numbers, as shown in Fig. \ref{Fig:16}(a). The interface lengths at Mach number $\text{M}_{s}=1.21$, 1.7, and 2.1 reach a minimum value at around $\tau=3$, $4$, and $10$, respectively. As indicated in Fig. \ref{Fig:16}(a), the upstream interface becomes flattered  at this moment. Later, after the compression phase, the variation of length with time is found almost constant at $\text{M}_{s}=1.21$, while these variations of length with time tends to increase at higher Mach numbers $\text{M}_{s}=1.7$, and 2.1 due to the enhanced rolled-up vortices, as seen in Fig. \ref{Fig:6}. Interestingly, the interface height increases constantly due to the continuous rotation of the vortex pair until the reflected shock waves strike the interface and reduce the interface height growth rate. Figure \ref{Fig:16}(b) shows that higher Mach number $(\text{M}_{s}=2.1)$ gives the maximum interface height, while lower Mach number $(\text{M}_{s}=1.21)$ produces the minimum interface height.

\section{Concluding remarks}
\label{Sec.5}

The Richtmyer-Meshkov (RM) instability has long been a fascinating subject due to its fundamental significance in scientific research, as well as its crucial role in engineering applications. In this work, the contribution of different incident shock Mach numbers ($\text{M}_{s}=1.21,1.7$ and 2.1) on the evolution of the RM instability in a shock-accelerated square light bubble is investigated numerically. A two-dimensional system of unsteady compressible Navier--Stokes--Fourier equations are solved by using an explicit mixed-type modal discontinuous Galerkin method with uniform meshes. The numerical results are compared to available experimental results for the validation study and are found to be in good agreement.

The numerical results reveal that the shock Mach numbers play a significant role in describing the RM instability during the interaction between a planar shock wave and a light bubble. The effects of Mach numbers result in a substantial change in the flow morphology with complex wave patterns, vortex creation, vorticity generation, and bubble deformation. In contrast to low Mach numbers, high Mach numbers produce larger rolled-up vortex chains, larger inward jet formation, and a stronger mixing zone with larger expansion. At high Mach numbers, the bubble deforms differently, and the reflected shock wave promotes a more complicated deformation of the bubble. Additionally, larger distortions of the bubble occur at the early time instants for higher Mach numbers.

A detailed study of the Mach numbers effects are investigated through vorticity generation and evolutions of enstrophy as well as dissipation rate. It is interesting to observe that vorticity plays a significant role to describe essential features in the study of the shock-accelerated bubble. It is found that the vorticity in the bubble region is enhanced with the increment of shock Mach number, especially for the period of the incident and reflected shock waves impinging on the bubbles. The results reveal that the viscous production term has the most dominant mechanism, followed by baroclinic vorticity and dilatational vorticity production terms in the interaction for strong shock waves. 
In addition, a significant increase in enstrophy and the dissipation rate is also found with increasing the shock Mach numbers. Finally, the time variations of the shock trajectories and interface features are investigated.

This study focuses on the contribution of Mach number on the evolution of the RM instability in a shock-accelerated square light bubble. It is also expected that the effect of Mach number will significantly affect the development of the RM instability in shock-accelerated heavy gas polygonal bubbles with different interface shapes. In this context, the present work will be extended in the future to report the results of our investigation into these problems.

\begin{acknowledgments}
The author would like to acknowledge the financial support of the NAP-SUG grant program funded by the Nanyang Technological University, Singapore.
The author is indebted to anonymous reviewers for their suggestions and remarks in light of which he could improve the article.
\end{acknowledgments}



\begin{thebibliography}{100}
\bibitem{Richtmyer1960taylor}
R. D. Richtmyer, Taylor instability in shock acceleration of compressible fluids, Commun. Pure. Appl. Math. 13, 297 (1960).

\bibitem{Meshkov1969Instability}
E. E. Meshkov, Instability of the interface of two gases accelerated by a shock wave, Fluid Dyn. 4, 101 (1969).

\bibitem{taylor1950instability}
G. I. Taylor, The instability of liquid surfaces when accelerated in a direction perpendicular to their planes. I, Proc. R. Soc. Lond. A 201, 192 (1950).

\bibitem{zabusky1999vortex}
N. J. Zabusky, Vortex paradigm for accelerated inhomogeneous flows: Visiometrics for the Rayleigh--Taylor and Richtmyer--Meshkov environments, Annu. Rev. Fluid Mech. 31, 495 (1999).

\bibitem{holmes1999richtmyer}
R. L. Holmes, G. Dimonte, B. Fryxell, M. L. Gittings, J. W. Grove, M. Schneider, D. H. Sharp, A. L. Velikovich, R. P. Weaver, Q. Zhang, Richtmyer--Meshkov instability growth: Experiment, simulation and theory, J. Fluid Mech. 389, 55 (1999).

\bibitem{brouillette2002richtmyer}
M. Brouillette, The Richtmyer--Meshkov instability, Annu. Rev. Fluid Mech. 34, 445 (2002).

\bibitem{ranjan2011shock}
D. Ranjan, J. Oakley, R. Bonazza, Shock-bubble interactions, Annu. Rev. Fluid Mech. 43, 117 (2011).

\bibitem{luo2014experimental}
X. Luo, Z. Zhai, T. Si, J. Yang, Experimental study on the interfacial instability induced by shock waves, Adv. Mech. 44, 260 (2014).

\bibitem{zhou2017rayleighI}
Y. Zhou, Rayleigh--Taylor and Richtmyer--Meshkov instability induced flow, turbulence, and mixing. I, Phys. Rep. 720–722, 1 (2017).

\bibitem{zhou2017rayleighII}
Y. Zhou, Rayleigh--Taylor and Richtmyer--Meshkov instability induced flow, turbulence, and mixing. II, Phys. Rep. 723, 1 (2017).


\bibitem{markstein1957shock}
G. H. Markstein, A shock-tube study of flame front-pressure wave interaction, in 6th International Symposium on Combustion (Elsevier, Amsterdam, 1957), Vol. 6, p. 387.

\bibitem{rudinger1960behaviour}
G. Rudinger, L. M. Somers, Behavior of small regions of different gases carried in accelerated gas flows, J. Fluid Mech. 7, 161 (1960).

\bibitem{haas1987interaction}
J. F. Haas, B. Sturtevant, Interaction of weak shock waves with cylindrical and spherical gas inhomogeneities, J. Fluid Mech. 181, 41(1987).

\bibitem{layes2005experimental}
G. Layes, G. Jourdan, L. Houas, Experimental investigation of the shock wave interaction with a spherical gas inhomogeneity,  Phys. Fluids 17, 028103 (2005).

\bibitem{layes2009experimental}
G. Layes, G. Jourdan, L. Houas, Experimental study on a plane shock wave accelerating a gas bubble, Phys. Fluids 21, 074102 (2009).

\bibitem{kumar2005stretching}
S. Kumar, G. Orlicz, C. Tomkins, C. Goodenough, K. Prestridge, P. Vorobieff, Stretching of material lines in shock-accelerated gaseous flows, Phys. Fluids 17, 082107 (2005).

\bibitem{ranjan2007experimental}
D. Ranjan, J. H. J. Niederhaus, B. Motl, M. H. Anderson, J. Oakley, R. Bonazza, Experimental investigation of primary and secondary features in high-Mach-number shock-bubble interaction, Phys. Rev. Lett. 98, 024502 (2007).

\bibitem{ranjan2008shock}
D. Ranjan, J. H. J. Niederhaus, J. G. Oakley, M. H. Anderson, R. Bonazza, J. A. Greenough, Shock-bubble interactions: Features of divergent shock--refraction geometry observed in experiments and simulations, Phys. Fluids 20, 036101 (2008).


\bibitem{haehn2010experimental}
N. Haehn, D. Ranjan, C. Weber, J. G. Oakley, M. H. Anderson, R. Bonazza, Experimental investigation of a twice--shocked spherical density inhomogeneity, Phys. Scr. 2010, 014067 (2010).

\bibitem{haehn2012experimental}
N. Haehn, C. Weber, J. Oakley, M. H. Anderson, D. Ranjan, R. Bonazza, Experimental study of the shock--bubble interaction with reshock, Shock Waves 22, 47--56 (2012).

\bibitem{si2012experimental}
T. Si, Z. Zhai, J. Yang, X. Luo, Experimental investigation of reshocked spherical gas interfaces, Phys. Fluids 24, 054101 (2012).


\bibitem{picone1988vorticity}
J. M. Picone, J. P. Boris, Vorticity generation by shock propagation through bubbles in a gas, J. Fluid Mech. 189, 23 (1988).

\bibitem{quirk1996dynamics}
J. J. Quirk, S. Karni, On the dynamics of a shock-bubble interaction, J. Fluid Mech. 318, 129 (1996).

\bibitem{zabusky1998shock}
N. J. Zabusky, S. M. Zeng, Shock cavity implosion morphologies and vortical projectile generation in axisymmetric shock-spherical fast/slow bubble interactions, J. Fluid Mech. 362, 327–346 (1998).

\bibitem{bagabir2001mach}
A. Bagabir, D. Drikakis, Mach number effects on shock-bubble interaction, Shock Waves 11(3), 209 (2001).

\bibitem{giordano2006richtmyer}
J. Giordano, Y. Burtschell, Richtmyer--Meshkov instability induced by shock-bubble interaction: Numerical and analytical studies with experimental validation, Phys. Fluids 18, 036102 (2006).

\bibitem{Niederhaus2008}
J. H. J. Niederhaus, J. A. Greenough, J. G. Oakley, D. Ranjan, M. H. Anderson, R. A. Bonazza, A computational parameter study for the three-dimensional shock-bubble interaction, J. Fluid Mech.
594, 85 (2008).

\bibitem{zhu2018numerical}
Y. Zhu, Z. Yang, Z. Pan, P. Zhang, J. Pan, Numerical investigation of shock-$\text{SF}_6$ bubble interaction with different Mach numbers, Comput. Fluids 177, 78 (2018).


\bibitem{rybakin2014supersonic}
B. Rybakin, V. Goryachev, The supersonic shock wave interaction with low-density gas bubble, Acta Astronaut. 94(2), 749 (2014).

\bibitem{singh2021nonequilibrium}
S. Singh, M. Battiato, Behavior of a shock accelerated heavy cylindrical bubble under nonequilibrium conditions of diatomic and polyatomic gases, Phys. Rev. Fluids 6, 044001 (2021).

\bibitem{singh2021bulk}
S. Singh, M. Battiato, R. S. Myong, Impact of the bulk viscosity on flow morphology of shock-bubble interaction in diatomic and polyatomic gases, Phys. Fluids 33, 066103 (2021) 


\bibitem{zhai2014interaction}
Z. Zhai, M. Wang, T. Si, X. Luo, On the interaction of a planar shock with a light polygonal interface, J. Fluid Mech. 757, 800 (2014). 

\bibitem{luo2015interaction}
X. Luo, M. Wang, T. Si, Z. Zhai, On the interaction of a planar shock with an $\text{SF}_{6}$ polygon, J. Fluid Mech. 773, 366(2015). 

\bibitem{igra2018numerical}
D. Igra, O. Igra, Numerical investigation of the interaction between a planar shock wave with square and triangular bubbles containing different gases, Phys. Fluids 30, 056104 (2018).

\bibitem{igra2020shock}
D. Igra, O. Igra, Shock wave interaction with a polygonal bubble containing two different gases, a numerical investigation, J. Fluid Mech. 889, 1(2020). 

\bibitem{fan2019numerical}
E. Fan, B. Guan, C.-Y. Wen, H. Shen, Numerical study on the jet formation of simple-geometry heavy gas inhomogeneities, Phys. Fluids 31, 026103 (2019).

\bibitem{singh2020role}
S. Singh, Role of Atwood number on flow morphology of a planar shock-accelerated square bubble: A numerical study, Phys. Fluids 32, 126112 (2020).

\bibitem{singh2021IJHMT}
S. Singh, Numerical investigation of thermal non-equilibrium effects of diatomic and polyatomic gases on the shock-accelerated square light bubble using a mixed-type modal discontinuous Galerkin method, Int. J. Heat Mass Transf. 179, 121708 (2021).



\bibitem{Shankar2011Kawai}
S. K. Shankar, S. Kawai, S. K. Lele, Two-dimensional viscous flow simulation of a shock accelerated heavy gas cylinder, Phys. Fluids 23, 024102 (2011).	

\bibitem{Picone1988}
J. M. Picone, J. P. Boris, Vorticity generation by shock propagation through bubbles in a gas, J. Fluid Mech. 189, 23–51 (1988).

\bibitem{Samtaney1994}
R. Samtaney, N. J. Zabusky, Circulation deposition on shock-accelerated planar and curved
density-stratified interfaces: models and scaling laws, J. Fluid Mech. 269, 45–78 (1994).

\bibitem{Latini2020RMI}
M. Latini, O. Schilling, A comparison of two-and three-dimensional single-mode reshocked Richtmyer-Meshkov instability growth, Physica D 401, 132201 (2020).

\bibitem{Bird194}
G. Bird, Molecular Gas Dynamics and the Direct Simulation of Gas Flows, Clarendon,Oxford (1994).  


\bibitem{Xiao2014Myong}
N. T. P. Le, H. Xiao, R. S. Myong, A triangular discontinuous Galerkin method for non--Newtonian implicit constitutive models of rarefied and microscale gases, J. Comput. Phys. 273, 160 (2014).

\bibitem{Prince2018}
L. Prince Raj, S. Singh, A. Karchani, R. S. Myong, A super-parallel mixed explicit discontinuous Galerkin method for the second-order Boltzmann-based constitutive models of rarefied and microscale gases, Comput. Phys. 157, 146 (2017).

\bibitem{hill2006large}
D. J. Hill, C. Pantano, D. I. Pullin, Large-eddy simulation and multiscale modelling of a Richtmyer--Meshkov instability with reshock, J. Fluid Mech. 557, 29--61 (2006).

\bibitem{thornber2010influence}
B. Thornber, D. Drikakis, D. L. Youngs, R. J. R. Williams, The influence of initial conditions on turbulent mixing due to Richtmyer--Meshkov instability, J. Fluid Mech. 654, 99--139 (2010).

\bibitem{hahn2011richtmyer}
M. Hahn, D. Drikakis, D. L. Youngs, R. J. R. Williams, Richtmyer--Meshkov turbulent mixing arising from an inclined material interface with realistic surface perturbations and reshocked flow, Phys. Fluids 23, 046101 (2011).

\bibitem{balasubramanian2012experimental}
S. Balasubramanian, G. C. Orlicz, K. P. Prestridge, B. J. Balakumar, Experimental study of initial condition dependence on Richtmyer--Meshkov instability in the presence of reshock, Phys. Fluids 24, 034103 (2012).

\bibitem{mansoor2020effect}
M. M. Mansoor, S. M. Dalton, A. A. Martinez, T. Desjardins, J. J. Charonko, K. P. Prestridge, The effect of initial conditions on mixing transition of the Richtmyer--Meshkov instability, J. Fluid Mech. 904, A3 (2020).

\bibitem{kundu2019high}
A. Kundu, S. De, High resolution numerical simulation of a shock-accelerated refrigerant-22 bubble, J. Comput. Physics 135, 260 (1997).

\bibitem{harten1997high}
A. Harten, High resolution schemes for hyperbolic conservation laws, Comput. Fluids 193, 104289 (2019).

\bibitem{karniadakis2013spectral}
G. Karniadakis, S. Sherwin, Spectral/hp element methods for computational fluid dynamics, Oxford University Press (2013).


\bibitem{bagabir2004numerical}
A. Bagabir, D. Drikakis, Numerical experiments using high-resolution schemes for unsteady, inviscid, compressible flows, Comput. Methods Appl. Mech. Engrg. 193, 4675 (2004).

\bibitem{mosedale2007assessment}
A. Mosedale, D. Drikakis, ssessment of very high order of accuracy in implicit LES models, J. Fluids Eng. 129, 1497 (2007).

\bibitem{Cockburn1998Local}
B. Cockburn, C.-W. Shu, The local discontinuous Galerkin method for time dependent
convection-diffusion systems, SIAM J. Numer. Anal. 35, 2440 (1998).

\bibitem{Cockburn1998DGV}
B. Cockburn, C.-W. Shu, The Runge-Kutta discontinuous Galerkin method for conservation laws V: Multidimensional systems, J. Comput. Phys. 141, 199 (1998).

\bibitem{singh2017JCFE}
S. Singh, R. S. Myong, A computational study of bulk viscosity effects on shock-vortex
interaction using discontinuous Galerkin method, J. Comput. Fluids Eng. 22, 86 (2017).

\bibitem{Signh2018Thesis}
S. Singh, Development of a 3D discontinuous Galerkin method for the second-order Boltzmann-Curtiss based hydrodynamic models of diatomic and polyatomic gases, Ph.D. Thesis, Gyeongsang National University, South Korea (2018).

\bibitem{Singh2020IACM}
S. Singh, M. Battiato, Strongly out-of-equilibrium simulations for electron Boltzmann transport equation using explicit modal discontinuous Galerkin method, Int. J. Appl. Comput. Math. 6, 1 (2020).

\bibitem{Singh2020Material}
S. Singh, M. Battiato, Effect of Strong Electric Fields on Material Responses: The Bloch Oscillation Resonance in High Field Conductivities, Materials 13, 1070 (2020).

\bibitem{Chourushi2020Singh}
T. Chourushi, A. Rahimi, S. Singh, R. S. Myong, Computational simulations of near-continuum gas flow using Navier–Stokes-Fourier equations with slip and jump conditions based on the modal discontinuous Galerkin method, Adv. Aerodyn. 2, 653 (2020).

\bibitem{singh2018non}
S. Singh, A. Karchani, R. S. Myong, Non-equilibrium effects of diatomic and polyatomic gases on the shock-vortex interaction based on the second-order constitutive model of the Boltzmann-Curtiss equation, Phys. Fluids 30, 016109 (2018).

\bibitem{singh2020topology}
S. Singh, A. Karchani, K. Sharma, R. S. Myong, Topology of the second-order constitutive model based on the Boltzmann--Curtiss kinetic equation for diatomic and polyatomic gases, Phys. Fluids 32, 026104 (2020).

\bibitem{Singh2021CF}
S. Singh, M. Battiato, An explicit modal discontinuous Galerkin method for Boltzmann transport equation under electronic nonequilibrium conditions, Comput. Fluids 224, 104972 (2021).

\bibitem{Singh2021JCP}
S. Singh, A. Karchani, T. Chourushi, R. S. Myong, A three-dimensional modal discontinuous Galerkin method for second-order Boltzmann-Curtiss constitutive models of rarefied and microscale gas flows,  under review in J. Comput. Phys. (2021).

\bibitem{Roe1981JCP}
P. L. Roe, Approximate riemann solvers, parameter vectors and difference schemes, J. Comput. Phys. 43, 357(1981).


\bibitem{schilling2007physics}
O. Schilling, M. Latini, W. S. Don,  Physics of reshock and mixing in single--mode Richtmyer--Meshkov instability, Phys. Rev. E 76, 026319 (2007).

\bibitem{Marquina2003Mulet}
A. Marquina, P.  Mulet,  A flux-split algorithm applied to conservative models for multicomponent compressible flows, J. Comput. Phys. 185, 120 (2003).

\bibitem{smirnov2014hydrogen}
N. N. Smirnov, V. B. Betelin, R. M. Shagaliev, V. F. Nikitin, I. M. Belyakov, Y. N. Deryuguin, S. V. Aksenov, D. A. Korchazhkin, Hydrogen fuel rocket engines simulation using LOGOS code, Int. J. Hydrog. Energy 39, 10748 (2014).

\bibitem{smirnov2015accumulation}
N. N. Smirnov, V. B. Betelin, V. F. Nikitin, L. I. Stamov, D. I. Altoukhov, Accumulation of errors in numerical simulations of chemically reacting gas dynamics, Acta Astronaut. 117, 338 (2015).

\bibitem{foias2001navier}
C. Foias, O. Manley, R. Rosa, R. Temam, Navier-Stokes equations and turbulences, Cambridge University Press (2001).




\end{thebibliography}
\end{document}